\documentclass[aps,12pt,superscriptaddress,floatfix,nofootinbib,altaffilletter]{revtex4-1}
\usepackage{graphicx}
\usepackage[colorlinks=true,citecolor=darkred,urlcolor=darkred, pdfborder={0 0 0}]{hyperref}
\usepackage[utf8]{inputenc} 
\usepackage{amsmath,amssymb,amsbsy}
\usepackage[usenames,dvipsnames]{color}  %%%%%%% usenames,dvipsnames required to get BrickRed
\usepackage{float}
\usepackage{calc}
\usepackage{comment}
\usepackage{multirow}
\usepackage{slashed}
\usepackage[normalem]{ulem}
\usepackage{braket}
\usepackage{mathtools}
\usepackage{caption}
\usepackage{subcaption}
\usepackage{wrapfig}
\reversemarginpar
\usepackage{xcolor}

\usepackage{paralist}
\usepackage{comment}
\usepackage[T1]{fontenc}

\definecolor{darkred}{rgb}{0.6,0,0}

\usepackage{enumitem}
\usepackage{cancel}

\def\znbb {$0\nu\beta\beta$ }

\def\Z2{$\mathcal{Z_2}$}

\usepackage{ragged2e}
 \newcommand{\bl}[1]{{\color{magenta}  #1}}

\newcommand {\ignore}[1]{}

\usepackage{mathrsfs}
\def\SM{$\mathrm{SU(3)_c \otimes SU(2)_L \otimes U(1)_Y}$ }

\def\321{$\mathrm{SU(3) \otimes SU(2) \otimes U(1)}$ }

\definecolor{linkcolor}{rgb}{0,0,0.5}

\def\bl#1{\textcolor{blue}{#1}}

\newcommand{\AddrAHEP}{
  AHEP Group, Institut de F\'{i}sica Corpuscular --
  CSIC/Universitat de Val\`{e}ncia, Parc Cient\'ific de Paterna.\\
 C/ Catedr\'atico Jos\'e Beltr\'an, 2 E-46980 Paterna (Valencia) - SPAIN}
 
\newcommand{\AddrIISERB}{Department of Physics, Indian Institute of Science Education and Research - Bhopal, \\ 
Bhopal Bypass Road, Bhauri, Bhopal 462066, India}

\newcommand{\AddrCFTP}{Departamento de F\'{\i}sica and CFTP, Instituto Superior T\'ecnico, Universidade de Lisboa, Av. Rovisco Pais 1, 1049-001 Lisboa, Portugal}

\begin{document}

\title{\color{BrickRed} Large lepton number violation at colliders:\\
 predictions from the minimal linear seesaw mechanism
}
\author{Aditya Batra}\email{aditya.batra@tecnico.ulisboa.pt}
\affiliation{\AddrIISERB}
\affiliation{\AddrCFTP}
\author{ Praveen Bharadwaj}\email{praveen20@iiserb.ac.in}
\affiliation{\AddrIISERB}
\author{Sanjoy Mandal}\email{smandal@kias.re.kr}
\affiliation{Korea Institute for Advanced Study, Seoul 02455, Korea}
\author{Rahul Srivastava}\email{rahul@iiserb.ac.in}
\affiliation{\AddrIISERB}
\author{Jos\'{e} W. F. Valle}\email{valle@ific.uv.es}
\affiliation{\AddrAHEP}
%%%%%%%%%%%%%%%%%%%%%%%%%%
\begin{abstract} 
    Small neutrino masses can be sourced by a tiny vacuum expectation value of a leptophilic Higgs doublet, and mediated by Quasi-Dirac heavy neutrinos.
    In such simplest linear seesaw picture the neutrino mass mediators can be accessible to colliders.
  We describe novel charged Higgs and heavy neutrino production mechanisms that can be sizeable at $e^+ e^-$, $e^- \gamma$, $pp$, or muon colliders and discuss some of the associated signatures.
  The oscillation length of the heavy neutrino mediators is directly related to the light neutrino mass ordering.
  Moreover, lepton number violation can be large despite the smallness of neutrino masses, and may shed light on the Majorana nature of neutrinos and the significance of basic symmetries in weak interaction.
  %lepton number and lepton flavour non-conservation. }
  \end{abstract}
%%%%%%%%%%%%%%%%%%%%%%%%%%%%
\maketitle
\section{Introduction}\label{sec:introduction} 
The discovery of neutrino oscillations~\cite{Kajita:2016cak,McDonald:2016ixn} brought neutrinos to the spotlight in particle physics.
Non-zero neutrino masses constitute one of the most convincing proofs of new physics, required to account for the oscillation data~\cite{deSalas:2020pgw}. 
Although the SM lacks neutrino masses, Weinberg's dimension-five operator $\frac{1}{\Lambda}({L} \Phi)^2 $ 
(where $L$ is the lepton doublet and $\Phi$ is the SM Higgs doublet) breaks lepton number by two units~\cite{Weinberg:1980bf}.
Below the electroweak symmetry breaking scale, it leads to small Majorana neutrino masses.
The effective mass scale $\Lambda$ depends on the underlying dynamics. 

The popular ``seesaw-completion'' of Weinberg's operator postulates that neutrinos get mass from the exchange of heavy SM singlet mediators.
In its conventional ``high-scale'' implementation the type-I seesaw mediators are superheavy, and hence kinematically inaccessible. 
If one ``artificially'' takes these mediators to have low mass, they will be so weakly coupled in the SM gauge currents,
that they will not be copiously produced in collider setups. 
In fact, any direct manifestation of their existence other than those associated to the neutrino masses, e.g. oscillations and \znbb decay, is washed out.
This will preclude a direct experimental test of the validity of the seesaw picture. 

However, in its most general SM-based realization, the seesaw allows for any number of singlet mediators~\cite{Schechter:1980gr}. 
Indeed, the inverse~\cite{Mohapatra:1986bd,Gonzalez-Garcia:1988okv} and the linear seesaw~\cite{Akhmedov:1995ip,Akhmedov:1995vm,Malinsky:2005bi}
include a sequential pair of mediators, associated to each family. 
In contrast to high-scale seesaw, the heavy low-scale seesaw mediators may be produced at high-energy colliders~\cite{Dittmar:1989yg,Gonzalez-Garcia:1990sbd,AguilarSaavedra:2012fu,Das:2012ii,Deppisch:2013cya}.

Recently there have been several low-scale seesaw studies of collider signatures focusing on the issue of Lepton Number Violation (LNV)~\cite{Anamiati:2016uxp,Antusch:2017ebe,Drewes:2019byd,Fernandez-Martinez:2022gsu,Antusch:2022ceb,Antusch:2023nqd}.
Here, we examine this issue within the simplest variant of the linear seesaw mechanism. 
In addition to the pairs of lepton mediators, each carrying opposite lepton numbers, we add a leptophilic Higgs doublet, carrying two units of lepton number, and sourcing the small neutrino masses.
These are dynamically related to the tiny vacuum expectation value of this second doublet. 
This way, the heavy mediator neutrinos can lie in the TeV scale, while having sizeable Yukawa couplings. 
In contrast to most previous formulations~\cite{Akhmedov:1995ip,Akhmedov:1995vm,Malinsky:2005bi}, here we assume just the minimal \SM gauge structure. 
We show how the new leptophilic Higgs doublet can provide an efficient heavy neutrino production portal and large rates for Lepton Number Violation (LNV) at high energies. 
In this simple manner, the seesaw picture becomes testable at planned lepton colliders such as the ILC~\cite{Behnke:2013xla}, CLIC~\cite{CLIC:2018fvx}, FCC-ee~\cite{FCC:2018evy}
CEPC~\cite{CEPCStudyGroup:2018ghi} or a muon collider~\cite{Delahaye:2019omf}. 
These are still in the discussion phase, so our work should serve as motivation for the community to further consider the case for lepton colliders.

\section{Linear seesaw}\label{sec:model}
The simplest linear seesaw model supplements the SM Higgs doublet $\Phi$ with an extra lepton-number-carrying doublet $\chi_L$ that sources neutrino mass generation,
mediated by the lepton singlet pair $\nu^c$ and $S$ carrying opposite lepton numbers $L[S]=1=-L[\nu^c]$~\cite{Akhmedov:1995ip,Akhmedov:1995vm,Malinsky:2005bi}.   
By having just the minimal~\SM gauge structure with ungauged lepton number, broken only in the scalar sector, we avoid the existence of a physical Nambu-Goldstone boson,
and the associated restriction from LEP~\cite{Joshipura:1992hp}, as well as the stringent astrophysical limits from stellar cooling~\cite{Fontes:2019uld}.  
The scalar potential is given by 
%%%%%%%%%%%
\begin{align}
 V&=-\mu_\Phi^2 \Phi^{\dagger}\Phi - \mu_\chi^2 \chi_L^{\dagger}\chi_L+\lambda_1 (\Phi^{\dagger}\Phi)^2 + \lambda_2 (\chi_L^{\dagger}\chi_L)^2+\lambda_3 \chi_L^{\dagger}\chi_L \Phi^{\dagger}\Phi\nonumber \\
 & + \lambda_4 \chi_L^{\dagger}\Phi \Phi^{\dagger}\chi_L - \left(\mu_{12}^2 \Phi^{\dagger}\chi_L+ \text{h.c.} \right),
 \label{eq:potential}
\end{align}
%%%%%%%%%
For definiteness, we assume all parameters to be real. Notice that the small breaking of lepton number through the last bilinear term $\mu_{12}^2(\Phi^\dagger\chi_L + \text{h.c.})$ is explicit, but soft. 
This in turn induces a non-zero VEV for $\chi_L$, $v_\chi\approx \mu_{12}^2 v_\Phi/m_A^2$ where $m_A$ is the mass of pseudoscalar. A detailed analysis of the scalar sector can be found, e.g.,~in~\cite{Batra:2022arl}.

The relevant lepton-number-invariant Lagrangian for neutrino mass generation is 
\begin{equation}
  \label{eq:Yukawa}
  - \mathcal{L}_{\rm Yuk}= Y_{\nu}^{\alpha} L_\alpha^T C \nu^c \Phi 
  + M_R \nu^c C S + Y_{S}^{\alpha} L_\alpha^T C  S \chi_L+ \text{h.c.} 
\end{equation}
%%%%%%%%%%%
where the vector $\mathbf{Y}_{\nu,S}=(Y_{\nu,S}^e, Y_{\nu,S}^\mu, Y_{\nu,S}^\tau)^T$ encodes the neutrino Yukawa couplings of the leptophilic Higgs and $M_R$ is a Dirac-type mass term~\footnote{We use boldface font to represent matrices in flavor space, suppressing the generational index.}.
Note that, instead of the sequential model with three species of each neutrino type, here we focus on the minimal one containing just one species of singlets $\nu^c$ and $S$, distinguished by their lepton numbers.
Once lepton number is violated, we have the following linear seesaw mass matrix in the basis $\nu,\nu^c,S$ 
\begin{align}
\mathcal{M}_{\nu}=
 \begin{pmatrix}
  0 & \mathbf{m}_D  & \mathbf{M_L}  \\
  \mathbf{m}_D^T & 0 & M_R \\
  \mathbf{M}_L^T &  M_R  &  0  \\
 \end{pmatrix},
 \label{eq:neutrino-mass-matrix}
\end{align}
where $\mathbf{m}_D=\frac{\mathbf{Y}_\nu v_{\Phi}}{\sqrt{2}}$, $\mathbf{M}_L=\frac{\mathbf{Y}_S v_{\chi}}{\sqrt{2}}$ with $v_\Phi$, $v_\chi$ denoting the vacuum expectation values~(VEVs) of the $\Phi$ and $\chi_L$ doublets. 
The two complex doublets lead to five physical scalars. Besides the ``SM-like'' Higgs boson $h$ there are heavier states $H$ (CP-even) and $A$ (CP-odd) 
and the charged scalar $H^\pm$ that plays a key role as portal.
The VEV of the $\chi_L$ doublet is induced by an explicit lepton number violating term in the scalar sector. 
Hence it can be naturally small in t'Hooft's sense, as required to account for the small neutrino masses.
Following the standard seesaw diagonalization procedure~\cite{Schechter:1981cv} with $M_R\gg \mathbf{m}_D\gg \mathbf{M}_L$ in Eq.~\eqref{eq:neutrino-mass-matrix},
we get the effective light neutrino mass matrix as
\begin{equation}\label{lin}
m_{\nu}=\frac{\mathbf{m}_D\mathbf{M}_L^T+\mathbf{M}_L \mathbf{m}_D^T}{M_R}. 
\end{equation}
In contrast with conventional type-I seesaw, the matrix $m_{\nu}$ scales linearly with the Dirac term $\mathbf{m}_D$, hence the name linear seesaw mechanism.
The limit $\mathbf{M}_L\to 0$ leads to three massless neutrinos as in the SM, plus one heavy Dirac lepton. This constitutes the template  
for low-scale seesaw schemes, including also the inverse seesaw~\cite{Mohapatra:1986bd,Gonzalez-Garcia:1988okv}. 
A remarkable feature of these schemes is that they violate lepton flavour and CP symmetries even in the absence of neutrino
mass~\cite{Valle:1987gv,Bernabeu:1987gr,Langacker:1988up,Branco:1989bn,Rius:1989gk}, elucidating the role of these
symmetries and suggesting that the associated charged lepton flavour violating processes can be sizeable. 
Likewise, the non-trivial structure of the weak currents
  also implies unitarity-violation effects in neutrino propagation~\cite{Valle:1987gv,Langacker:1988up,Escrihuela:2015wra,Escrihuela:2016ube,Miranda:2016wdr}. 

 We now turn on lepton number violation through a non-zero $\mathbf{M}_L$. This leads to small masses to the three Majorana eigenstates $\nu_i$, with $i=1,2,3$.
  The heavy ones $N_4$ and $N_5$ form one Quasi-Dirac fermion~\cite{Valle:1982yw} that can be accessible to high energy colliders, as we show below. 
In the limit of small induced VEV $v_\chi \to 0$ lepton number is restored, so the construction is natural in t'Hooft's sense. 
The above neutrino mass matrix can be diagonalised by a $5\times 5$ unitary matrix $\mathcal{U}$ as $\mathcal{U}^\dagger \mathcal{M}_{\nu} \mathcal{U}^*=\mathcal{M}_{\nu}^{\rm diag}$~\cite{Schechter:1980gr}, where $\mathcal{M}_{\nu}^{\rm diag}=\text{diag}(m_1, m_2, m_3, M_{N_4}, M_{N_5})$. The explicit form of the unitary matrix $\mathcal{U}$ is as follows,
\begin{eqnarray}\label{u-bdiag}
&\mathcal{U}\approx
\left(
\begin{array}{ccc}
U_{\rm } & -\frac{i}{\sqrt{2}}\mathbf{S} & \frac{1}{\sqrt{2}}\mathbf{S}\\
0 & \frac{i}{\sqrt{2}} & \frac{1}{\sqrt{2}}  \\
 -\mathbf{S}^{\dagger} & -\frac{i}{\sqrt{2}} & \frac{1}{\sqrt{2}} 
\end{array}
\right),  &
\end{eqnarray}
where the seesaw expansion parameter~\cite{Schechter:1981cv} 
\begin{equation*}  
    \textbf{S}=(\textbf{m}_D+\textbf{M}_L)/M_R.
\end{equation*}

Since $v_\chi$ is tiny the contribution from the term proportional to $Y_S$ is negligible, we can take 
\begin{align} 
\mathbf{S} \approx \mathbf{m}_D/M_R,
\end{align}
to characterize the light-heavy neutrino mixing in the weak interaction currents. 
\par Indeed one can show that, in the scenario with just one $\nu^c$ and one $S$, the Yukawa couplings $Y_\nu$ and $Y_S$ are fully determined. This can be seen by parameterizing $Y_\nu$ and $Y_S$ as,
%%%%%%%%%%%%%%%%%%%%%%%%%%%%%%%%%%%%%%%%%%%%%
\begin{eqnarray}
Y_\nu \equiv y_\nu \mathbf{u},  \;\;\;  {Y_S} \equiv y_S \mathbf{v}, \;\;\; 
\end{eqnarray}
%%%%%%%%%%%%%%%%%%%%%%%%%%%%%%%%%%
where $y_\nu$ and $y_S$ are real numbers and $\mathbf{u}$ and $\mathbf{v}$ are three complex vectors with unit norm. \\
%%%%%%%%%%%%%%%%%%%%%%%%%%%%%%%%%%%
For example, for the normal ordering case, the neutrino mass eigenvalues are:
\begin{align}
m_1 = 0 \,, \quad |m_2|= {y_\nu y_S v_\Phi v_\chi\over 2M_R}~(1-\rho) \,, \quad  |m_3|= {y_\nu y_S v_\Phi v_\chi\over 2M_R}~(1+\rho) \,, 
\end{align}
%%%%%
so that the measured ratio of the solar and atmospheric neutrino squared mass splittings fixes the parameter $\rho$:
\begin{align}
r \equiv {|\Delta m^2_{\text{solar}}|\over |\Delta m^2_{\text{atmos}}|} = {|\Delta m^2_{12}| \over |\Delta m^2_{23}|}
\;,\;\;\;\; \rho= \frac{\sqrt{1+r}-\sqrt{r}}{\sqrt{1+r} +\sqrt{r}} \,\,. 
\end{align}
This allows us to read off the columns of the lepton mixing matrix, as follows 
\begin{eqnarray}
{Y_\nu}_i = { y_\nu\over \sqrt{2}} \left( \sqrt{1+\rho}~U_{i3} + e^{i\frac{\pi}{2}} \sqrt{1-\rho} ~U_{i2}\right)\,, \\
{Y_S}_i = {y_S \over \sqrt{2}} \left( \sqrt{1+\rho}~U_{i3} - e^{i\frac{\pi}{2}} \sqrt{1-\rho} ~U_{i2}\right). 
\end{eqnarray} 
%%%%%%%%%%%%%%%%%%%%%%%%%%%%%%%%%%%%%%%%%%%%%%%%
%\begin{align}
%2.337\times 10^{-13}\lesssim\frac{y_\nu y_S v \sin\beta \cos\beta}{M_R}\lesssim 2.43\times 10^{-13}
%\end{align}
%

One can check explicitly that sizeable $Y_S$ values can be consistent with the oscillation data.

The heavy neutrino mass eigenvalues are obtained as
%%%%%%%%%
\begin{align}
M_{N_{4,5}}=M_R \left(1+\frac{1}{2}|\mathbf{S}|^2\right)\mp \frac{1}{2 M_R}\left(\mathbf{M}_L^\dagger \mathbf{m}_D + \mathbf{m}_D^\dagger \mathbf{M}_L\right).
\end{align}
%%%%%%%%%
The mass splitting of the heavy neutrinos is  $\Delta M=2|\mathbf{M}_L^*\mathbf{S}|$. 
Note that our minimal linear seesaw scenario is a “missing partner” seesaw~\cite{Schechter:1980gr}, similar to recent scoto-seesaw constructions~\cite{Rojas:2018wym,Mandal:2021yph}. The mass matrix $m_\nu$ has two nonzero eigenvalues 

\begin{align}
m_\nu^i=|\mathbf{M}_L| |\mathbf{S}|\mp |\mathbf{M}_L^*\mathbf{S}|,
\begin{cases}
i= 2,\,3 \text{ for {\bf NO}}\\
i= 1,\,2 \text{ for {\bf IO}}
\end{cases}
\end{align}
leading to improved detection prospects for \znbb decay~\cite{Dolinski:2019nrj}.
Moreover, the heavy neutrino mass splitting $\Delta M$ in our scenario is given directly in terms of the measured mass squared differences of the light neutrinos as~\cite{Antusch:2017ebe}
%%%%%%%%%
\begin{align}
\Delta M^{\rm {\bf NO}}=\Delta m_{32},~~~
\Delta M^{\rm {\bf IO}}=\Delta m_{21}.
\end{align}
%%%%%%%%%
This provides a very simple connection between the measured neutrino mass splittings in neutrino oscillations and the mass splitting between the two heavy neutrino mediators.

The charged current interaction will be modified due to light-heavy neutrino mixing, becoming a rectangular matrix~\cite{Schechter:1980gr}. The relevant block describing the small charged current interaction involving the heavy neutrino states takes the following form,
%%%%%
\begin{align}
-\mathcal{L}_{\rm cc}=\frac{g}{\sqrt{2}}S^\alpha \overline{\ell_\alpha} \gamma^\mu P_L (-\frac{i}{\sqrt{2}} N_4 + \frac{1}{\sqrt{2}} N_5) W_\mu + \text{h.c}
\end{align}
%%%%
Moreover, the relevant Yukawa interaction takes the following form,
%%%%%%%
\begin{align}
-\mathcal{L}_{\rm Yuk} & \approx c_\alpha \frac{Y_S^\alpha}{\sqrt{2}} \overline{\nu_\alpha} P_R (\frac{i}{\sqrt{2}} N_4+ \frac{1}{\sqrt{2}} N_5 ) H + i s_\beta \frac{Y_S^\alpha}{\sqrt{2}} \overline{\nu_\alpha} P_R (\frac{i}{\sqrt{2}} N_4+ \frac{1}{\sqrt{2}} N_5) A \\ \nonumber
&+ s_\beta Y_S^{\alpha} \overline{\ell_\alpha} P_R (\frac{i}{\sqrt{2}} N_4+ \frac{1}{\sqrt{2}} N_5 ) H^- + \text{h.c}
\end{align}
%%%%
This coupling will play a crucial role in our present work, and will make the linear seesaw distinct from other, more conventional, seesaw mechanisms.
Note that the two heavy states form a quasi-Dirac fermion, so one can define two heavy neutrino states in the above Lagrangian as $N=(-i N_4 + N_5)/\sqrt{2}$ and $\overline{N}=(i N_4 + N_5)/\sqrt{2}$. $N$ is always produced together with an anti-lepton $\ell^+$ and $\overline{N}$ is produced together a lepton $\ell^-$. This justifies calling them as $N$ and $\overline{N}$. We will see later that these definition is very helpful for describing heavy neutrino-antineutrino oscillations.

\par 
The new Yukawa interaction terms $Y_{\nu}^{\alpha} L_\alpha^T C \nu^c \Phi$ and $Y_{S}^{\alpha} L_\alpha^T C  S \chi_L$ are not only responsible for the generation of neutrino masses,  
but they also give rise to charged lepton flavour violating processes, such as $\ell_i\to\ell_j\gamma$.
Apart from the usual charged current contribution, there is a new leptophilic Higgs-mediated contribution.
Detailed predictions depend on the Yukawa coupling $\mathbf{Y}_\nu$ or $\mathbf{Y}_S$ and can be managed to lie below current experimental limits even for large  $\mathbf{Y}_\nu$ or $\mathbf{Y}_S$ Yukawa couplings.\\[-.4cm]

\section{Collider Phenomenology} 
\label{sec:coll-phen}

All scalars as well as fermion singlet mediators of the linear seesaw can naturally lie in the GeV-TeV scale, hence accessible to future collider experiments. 
There are two viable possibilities: the heavy neutrinos $N_i$ can be either lighter or heavier than the new scalars. 
We first discuss the novel production mechanisms for the heavy neutrinos and scalars, and then discuss their decay modes and resulting collider signatures.  

\subsection{Direct heavy neutrino production}  
\label{sec:direct-heavy-neutr}

Heavy neutrinos can be produced in $e^+e^-$ or $pp$ collisions in many
ways~\cite{Dittmar:1989yg,Gonzalez-Garcia:1990sbd,Atre:2009rg,AguilarSaavedra:2012fu,Das:2012ii,Dev:2013wba,Deppisch:2013cya,Banerjee:2015gca,ATLAS:2019kpx,CMS:2022fut}.
In $pp$ collisions, the most extensively studied production mechanisms are the charged and neutral current Drell-Yan processes, 
$pp\to W^{\pm *}\to\ell^{\pm} N$ and $pp\to Z^{*}\to \nu N$. 
In $e^+e^-$ collisions the heavy neutrino can be produced as $e^+e^-\to\nu N$ through $W$ and $Z$ mediated t and s-channel processes.
However, in these channels the heavy neutrino production cross-section is suppressed due to the small mixing $|\mathbf{S}|$ between the light and heavy neutrinos.

\par
\begin{figure}[!htbp]
	\includegraphics[width=0.20\linewidth]{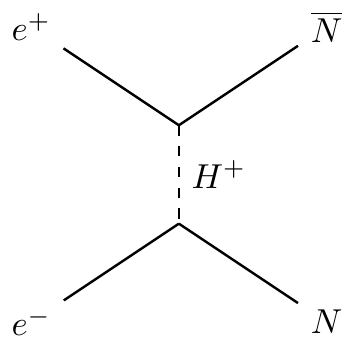}~~~~			
    	\includegraphics[width=0.22\linewidth]{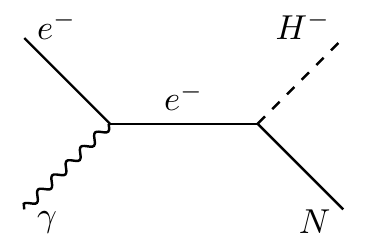}~~~~		
	\includegraphics[width=0.20\linewidth]{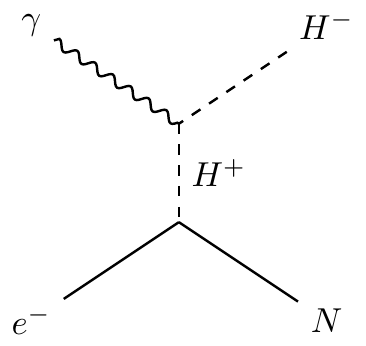}
	\caption{ 
          \footnotesize{Feynman diagrams for $N\overline{N}$ and $NH^{-}$ production in $e^+e^-$ or $e^-\gamma$ collisions, respectively.
           All these amplitudes are unsuppressed by light-heavy neutrino mixing $|S_{e}|$.}}
	\label{fig:HpN-production}
      \end{figure}

 Apart from the production through light-heavy neutrino mixing, in our linear seesaw scenario, the heavy neutrinos can also be pair-produced at $e^+e^-$ colliders
      through t-channel exchange of the leptophilic Higgs boson (first panel of Fig.~\ref{fig:HpN-production}). 
   \par Moreover, in the context of $e^+e^-$ colliders, the $e^+$ beam can be replaced by a backscattered photon, leading to an effective $e^-\gamma$
      collider that can have a rich physics potential~\cite{DeRoeck:2003cjp,DESY,Bechtel:2006mr,Telnov:2006cj}.
In this case, the heavy neutrinos can be produced in association with the charged scalar, as shown in the last two diagrams of Fig.~\ref{fig:HpN-production}. 
            
      The  $e^+e^-$ cross-section is proportional to $|Y_S^e|^4$, and hence can be large for large $|Y_S^e|$, whereas
the production cross-section contributions from the last two diagrams in Fig.~\ref{fig:HpN-production} are proportional to $|Y_S^e|^2$.  
Therefore, for relatively large Yukawa coupling $|Y_S^e|$, the $e^-\gamma\to N H^-$  process can be a promising way to produce heavy neutrinos.\\  
%%
%%
\iffalse
\begin{figure}[!htbp]
    	\includegraphics[width=0.8\linewidth]{eetoNNYs1.pdf}			
	\caption{
        \footnotesize{Heavy neutrino pair production cross-section versus the heavy neutrino mass $M_{N_i}$ at an $e^+e^-$ collider with center of mass energy $\sqrt{s}=3$ TeV, with $Y_S=\text{Diag}(1,1,1)$.}}
	\label{fig:eetoNN}
\end{figure}
%%
\begin{figure}[!htbp]
	\includegraphics[width=0.8\linewidth]{eatohmN.pdf}					
	\caption{
        \footnotesize{$NH^-$ production cross-section versus the heavy neutrino mass $M_{N_i}$ at an $e^-\gamma$ collider at center of mass energy $\sqrt{s}=3$ TeV, with $Y_S=\text{Diag}(1,1,1)$.}}
	\label{fig:eatohmN}
\end{figure}\fi

\subsection{Heavy neutrino decays} 
\label{sec:heavy-neutr-decays}

The $N_i$ mediators mix with the light neutrinos in the SM charged and neutral currents, as described in~\cite{Schechter:1980gr}.
They can decay through these charged and neutral current couplings into various SM final states, depending on their mass $M_{N_i}$.
For the case $M_{N_i}< m_W$ we have 3-body decays $N_i\to\ell_1\ell_2\nu$, $3\nu$ (purely leptonic) as well as semileptonic decays such as $\ell_1 u\bar{d}$ and $\nu_{\ell_1}q\bar{q}$.
On the other hand, for relatively large $M_{N_i}$, 2-body decay channels such as $\ell W$, $\nu_{\ell} Z$ and $\nu_{\ell} h$ start to dominate.  
The total $N_i$ decay width can be tiny for small mixing and relatively small $M_{N_i}$; hence these can be long-lived in that parameter region. 
However, this will not be the case for relatively large masses $M_{N_i} \geq \mathcal{O}(100\,\text{GeV})$ and sizeable
light-heavy neutrino mixing, which may be allowed in the linear seesaw. 
%%%%%%%%%%
\begin{figure}[!htbp]
\includegraphics[width=0.5\linewidth]{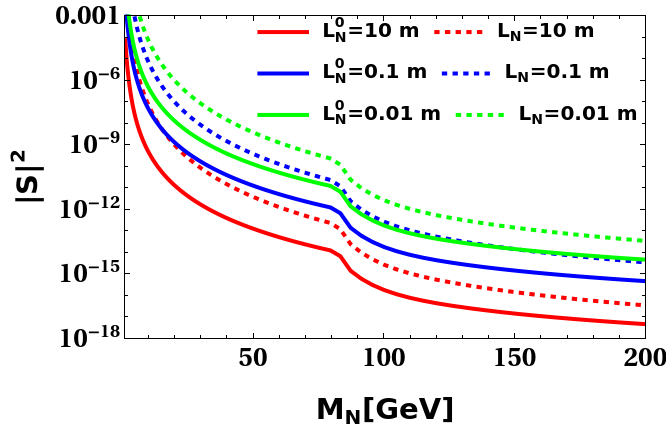}
\caption{\footnotesize{Contours of constant decay length of the heavy neutrinos $L_{N}^0=c\tau_N$ in the proper frame~(dashed lines) and in the laboratory frame~(solid lines). The decay length in the laboratory frame is given by $L_N=L_N^0\sqrt{\gamma^2-1}$ where $\gamma=E_N/M_N$ is the Lorentz factor for the specific process $e^+e^-\to N\overline{N}$.}}
\label{fig:mass-mixing}
\end{figure} 
%%%%%%%%%%

In Fig.~\ref{fig:mass-mixing}, we show the contours of constant decay length in the proper~(solid lines) and in the laboratory frame~(dashed lines). The decay length in the laboratory frame is given by $L_N=L_N^0\sqrt{\gamma^2-1}$, where $\gamma$ is the heavy neutrinos Lorentz factor for $e^+ e^-\to N\overline{N}$ with c.m. energy $\sqrt{s}=3$~TeV.
For $\gamma\gg 1$, the decay length in the laboratory frame will be enhanced.

Note that the discontinuity in $L_N^0$ and $L_N$ is due to the jump in $\Gamma_N$ around $M_N\sim 80$ GeV which comes from the threshold for gauge boson production $M_{W,Z}$. 
One sees that for sufficiently small mixing and relatively small $M_N$, the heavy neutrinos can have macroscopic lifetimes, so their decay occurs displaced from the primary vertex. 
This opens up the possibility of observing the oscillation patterns in the heavy neutrino decay spectra

\par Finally, for $M_{N_i}>m_{H^\pm},m_{H/A}$, small $v_\chi$ and large $|\mathbf{Y}_S|$, the decays $N_i\to\ell^\pm H^\mp$, $N_i\to\nu_\ell H/A$ dominate over those coming from light-heavy neutrino mixing.
As long as $|\mathbf{Y}_S|$ is large, the $N_i$ decay width is much larger than that coming just from light-heavy mixing, giving rise to prompt decays. \\

\subsection{Heavy neutrinos from charged Higgs bosons}
\label{sec:heavy-neutrinos-from}

\begin{figure}[!htbp]
	\includegraphics[width=0.22\linewidth]{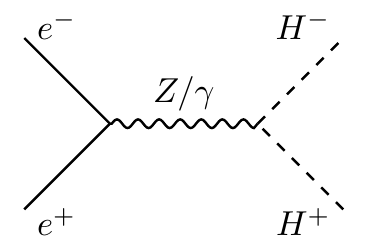}~~~~
	\includegraphics[width=0.20\linewidth]{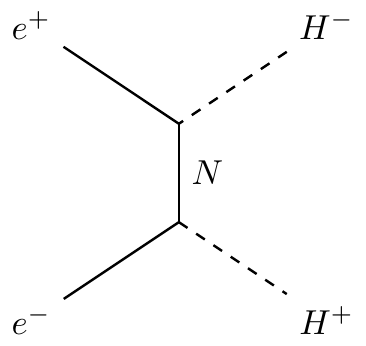}			
	\caption{\footnotesize{Feynman diagrams for charged Higgs boson pair production at an $e^+e^-$ collider.}}
	\label{fig:feynman-production}
\end{figure}

In the linear seesaw, the heavy neutrinos can also appear as the decay products of the charged scalar bosons produced in pairs at $e^+e^-$ colliders, see Fig.~\ref{fig:feynman-production}.

One finds that, for sufficiently large Yukawa coupling $|Y_S^e|\sim\mathcal{O}(1)$, the t-channel charged Higgs production process mediated by heavy neutrinos dominates over the s-channel process involving $\gamma/Z$ exchange.  
Thus for reasonably large $|Y_S^e|(\sim\mathcal{O}(1))$, the cross-section for this process is large enough ($\sim \mathcal{O}(100\, \text{fb})$) to be explored in future collider experiments.
%%
\iffalse
\begin{figure}[!htbp]		
	\includegraphics[width=0.8\linewidth]{eetohphp.pdf}
	%\includegraphics[width=1.\linewidth]{eetohphp2.pdf}
	\caption{
          \footnotesize{$e^+e^-\to H^+ H^-$ cross-section at $\sqrt{s}=3$ TeV with $Y_S$ chosen as Diag(1, 1, 1).
          The green line gives the s-channel contribution, while the others include both the s- and the t-channel
          for three mediator masses, $M_{N_i}=100$ GeV, 500 GeV, and 1 TeV. }}
	\label{fig:HpHm-t}
      \end{figure}\fi
      %%
As a result, we expect a large number of heavy neutrinos to be generated from the decay of these scalar particles produced at $e^+e^-$ colliders.  

\begin{center}
\bf $H/A$ and $H^\pm$ decays 
\end{center}
We now turn to the decays of the new scalar bosons, beyond the SM Higgs boson $h$.
Given the tiny value of the $U(1)$ breaking scale $v_\chi$, the neutral and charged scalars $H/A$ and $H^\pm$ are mainly composed of $\chi_L$.  
Hence couplings such as $(H/A)W^+W^-$, $(H/A)ZZ$, $(H/A)\ell\overline{\ell}$, $(H/A)q\overline{q}$, $AhZ$, $H^\pm q\overline{q'}$ and $H^\pm W^\mp h$ are all suppressed by $\mathcal{O}(v_\chi/v)$.
This makes the collider phenomenology quite different from the vanilla two-Higgs-doublet-model expectations.  
When the scalars are heavier than the neutrino mediators, the charged Higgs bosons dominantly decay into charged leptons $\ell^\pm$ and on-shell heavy neutrinos through the Yukawa interaction as
\begin{align}
\Gamma\left(H^+(H^-)\to \ell^+ N(\ell^-\overline{N})\right)\approx \frac{|Y_S^{\ell}|^2\sin^2\beta}{16\pi}m_{H^\pm}\Big(1-\frac{M_{N}^2}{m_{H^\pm}^2}\Big)^2,
\label{eq:HpDW}
\end{align}
whereas, the neutral Higgs bosons dominantly decay into light neutrinos $\nu_\ell$ and on-shell heavy neutrinos as  
\begin{align}
&\Gamma\left(H\to\nu_\ell N/\overline{N}\right)\approx \frac{|Y_S^{\ell}|^2\cos^2\alpha}{16\pi}m_H\Big(1-\frac{M_{N}^2}{m_{H}^2}\Big)^2,\label{eq:HDW}\\
& \Gamma(A\to\nu_\ell N/\overline{N})\approx \frac{|Y_S^{\ell}|^2 \sin^2\beta}{16\pi}m_{A}\Big(1-\frac{M_{N}^2}{m_{A}^2}\Big)^2,
\label{eq:ADW}
\end{align}
where, following the standard notation of the Two-Higgs-Doublet-Models, we have defined $\tan\beta =\frac{v_\Phi}{v_\chi}$ and denoted by $\alpha$ the mixing angle between the two CP even scalars.
As the decay widths are proportional to $|Y_S^{\ell}|$, they will be large for relatively large Yukawa coupling $|Y_S^{\ell}|$, favouring prompt decays. 
Note that for the opposite case of small $|Y_S^{\ell}|$ and relatively large $v_\chi$, the small scalar-sector mixing will be important and $H^\pm,H/A$ will dominantly decay into quarks or gauge bosons.
%%% 
\section{Lepton number violating signatures at colliders} 
We have seen that, at lepton colliders, production processes such as $e^+e^-\to N \overline{N}$, $e^-\gamma\to N H^-$ and $e^+ e^-\to H^+ H^-$ can be large, due to the large allowed values of the Yukawa coupling $|Y_S^e|$. 
This leads to the decay chain $H^+(H^-)\to\ell^+ N(\ell^- \overline{N})$, $N(\overline{N})\to\ell^- jj(\ell^+ j j)$~
 (note that if $m_{H^\pm}>M_{N}$ the dominant charged Higgs decay mode is $H^+(H^-)\to \ell^+ N~(\ell^-\overline{N})$ with heavy neutrinos decaying to SM final states via the light-heavy mixing),
giving rise to the following Lepton Number conserving (LNC) and violating (LNV) signatures:
\begin{align}
& e^+ e^- \to N \overline{N} \to \ell^\pm \ell^\pm 4j~(\text{LNV})/\ell^\pm \ell^\mp 4j~(\text{LNC}) \\
& e^- \gamma \to N H^- \to \ell^- \ell^\pm \ell^\pm 4j~(\text{LNV})/\ell^- \ell^\pm \ell^\mp 4j~(\text{LNC})\\
& e^+e^-\to H^+ H^- \to \ell^+ \ell^- \ell^\pm \ell^\pm 4j~(\text{LNV})/\ell^+ \ell^- \ell^\pm \ell^\mp 4j~(\text{LNC}).
\end{align}
%%%%%%%%%%%%%%%%%%%%%%%%%%%%%%%%
Note that within the linear seesaw with softly broken lepton number, one might except that LNV processes are severely suppressed. 
This follows from the fact that a quasi-Dirac fermion can be thought of as two fermions of opposite CP-phase and small Majorana mass-splitting ($\Delta M \sim \Delta m_\nu $)~\cite{Valle:1982yw}. 
In processes involving the virtual propagation of quasi-Dirac neutrinos, such as neutrinoless double beta decay ($0\nu 2\beta$), one has cancellations that suppress the LNV rates~\cite{Valle:1982yw}.
However, LNV processes producing quasi-Dirac fermions as real particles need not be suppressed. This can happen at collider energies~\cite{Anamiati:2016uxp,Antusch:2022ceb}.  
In this case oscillations can occur between the members of the quasi-Dirac pair, preventing the cancellations that would otherwise suppress the LNV signal. Assuming this oscillation is effective,~%($\sigma(\text{LNV})\approx \sigma(\text{LNC})$ 
which holds for an ideal detector as long as $\Delta M\gg \Gamma_N$ (see Sec.~\ref{sec:oscillation}), we now discuss the cross sections for all possible LNV final states coming from the production channels $e^+ e^- \to N \overline{N},\, e^- \gamma \to N H^-$ and $e^+e^-\to H^+ H^-$. In Sec.~\ref{sec:oscillation}, we discuss in detail the impact of heavy neutrino-antineutrino oscillations on the LNV rates, taking into account the finite detector size.
%%%%%%%%%
 \begin{figure*}[htbp]
	\includegraphics[width=0.32\linewidth]{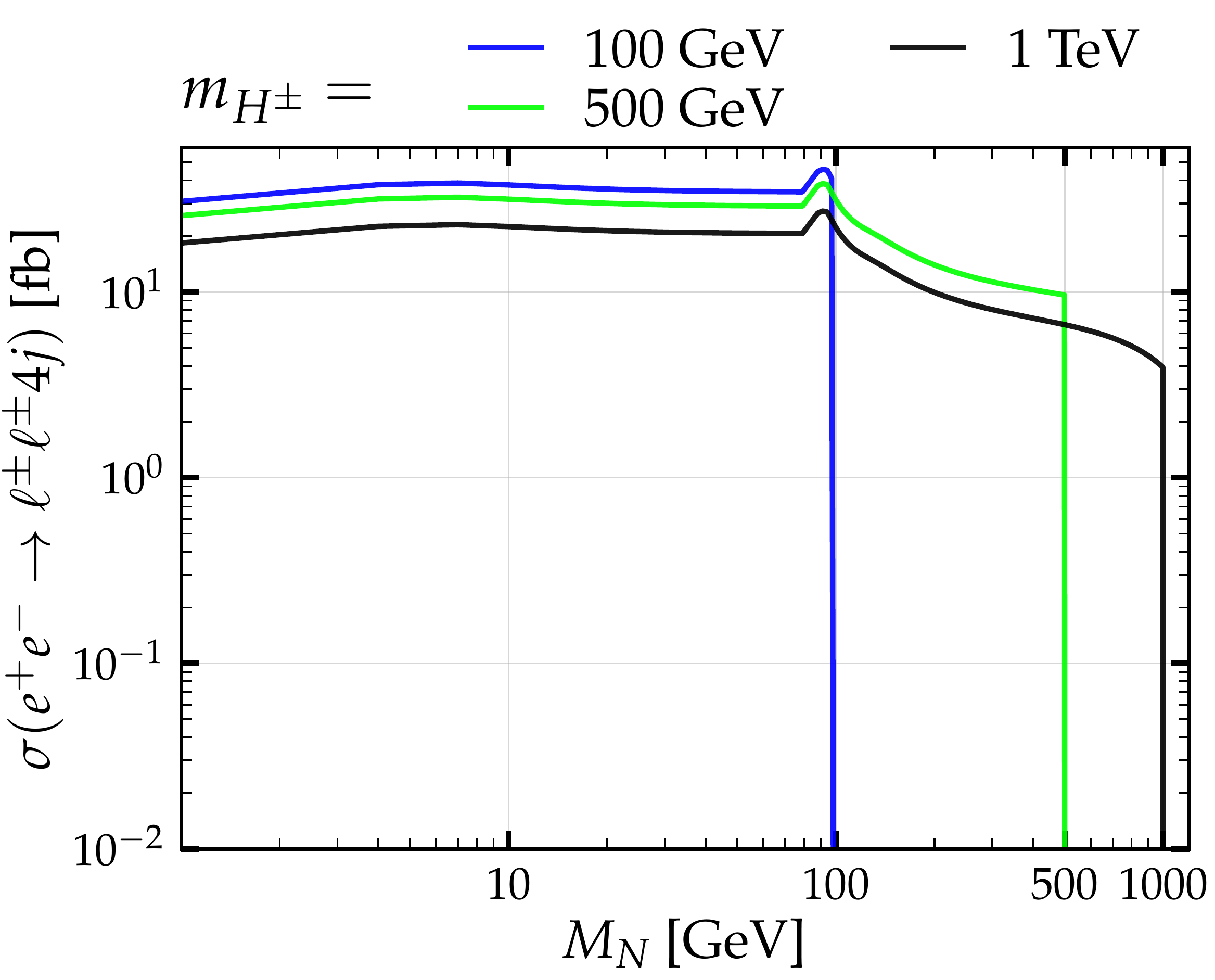}~~~	
	\includegraphics[width=0.32\linewidth]{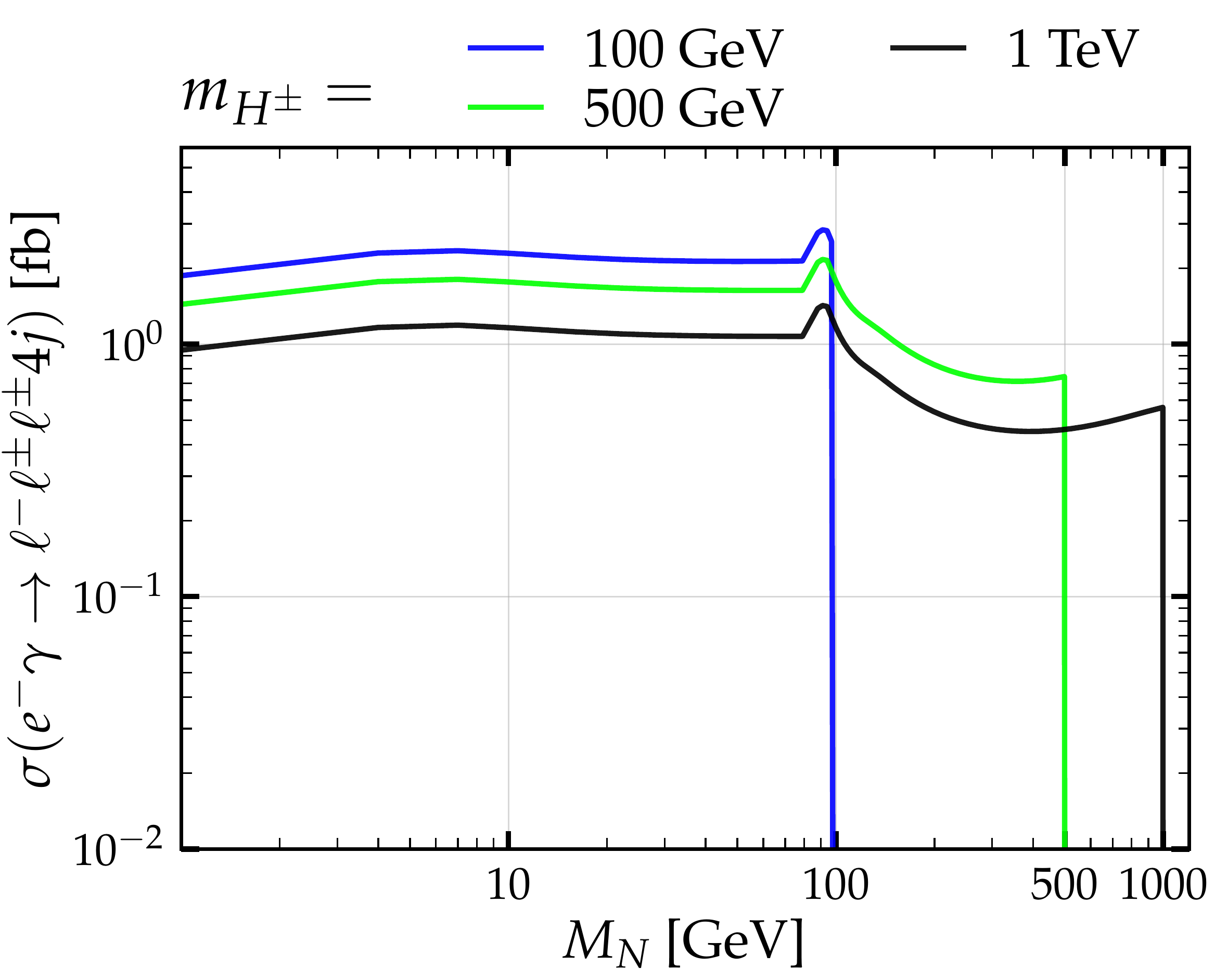}~~~
	\includegraphics[width=0.32\linewidth]{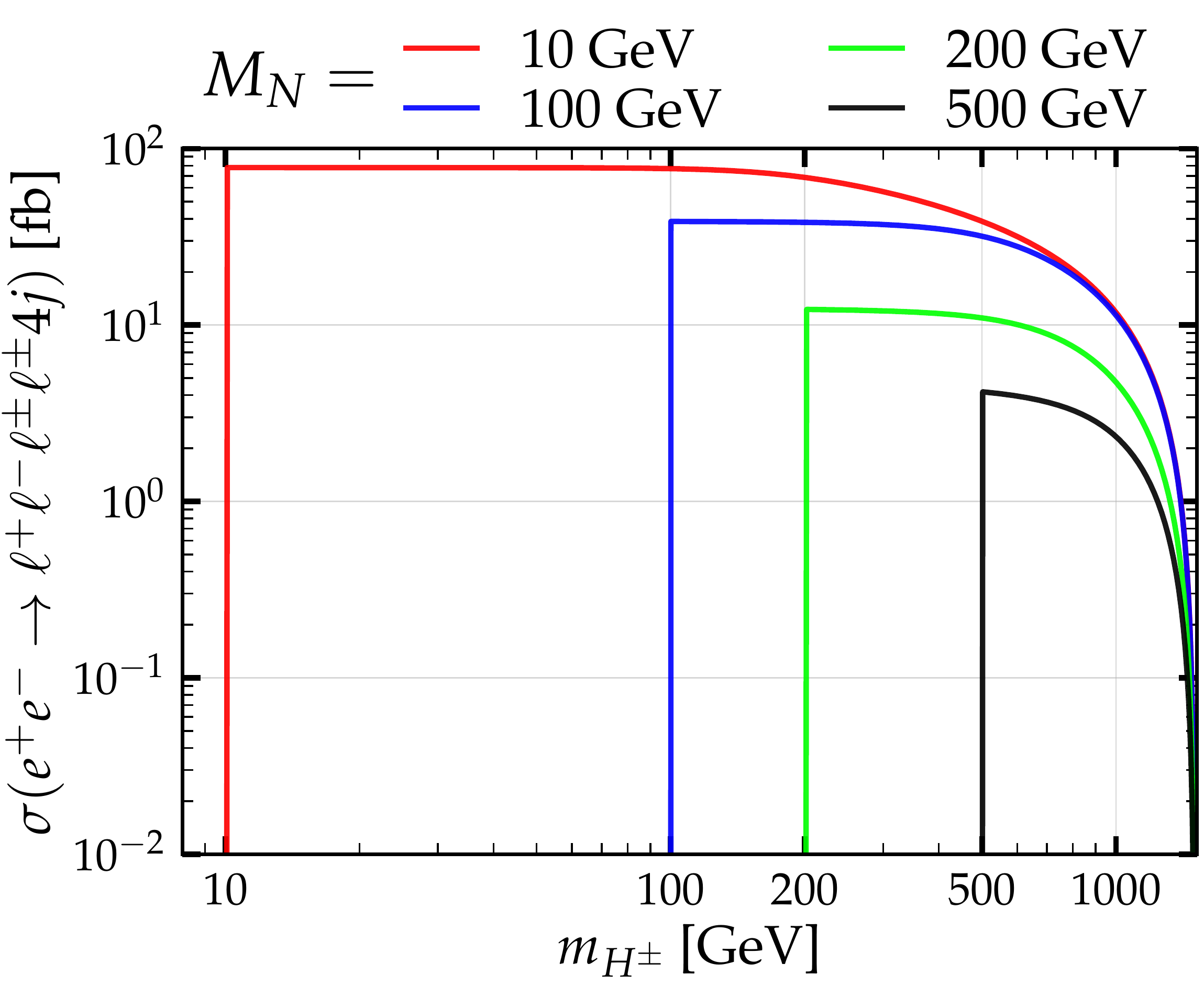}		
         \caption{
           \footnotesize{LNV cross-sections at $\sqrt{s}=$ 3~TeV center-of-mass energy for $e^+e^-\to N\overline{N}\to$ $\ell^\pm \ell^\pm 4j$~(left panel),~~$e^-\gamma\to H^-N\to \ell^- \ell^\pm \ell^\pm 4j$~(middle) 
           and for $e^+ e^-\to H^+H^-\to$ $\ell^+\ell^-\ell^\pm \ell^\pm 4j$ (right panel).
           For the Yukawa coupling we assume $\mathbf{Y}_S=(1,1,1)$. Note that all these rates are unsuppressed by light-heavy neutrino mixing $|S_{\ell}|$.}}
 	\label{fig:CSeetoee4j}
 \end{figure*}
\begin{figure*}[htbp]
	\includegraphics[width=0.32\linewidth]{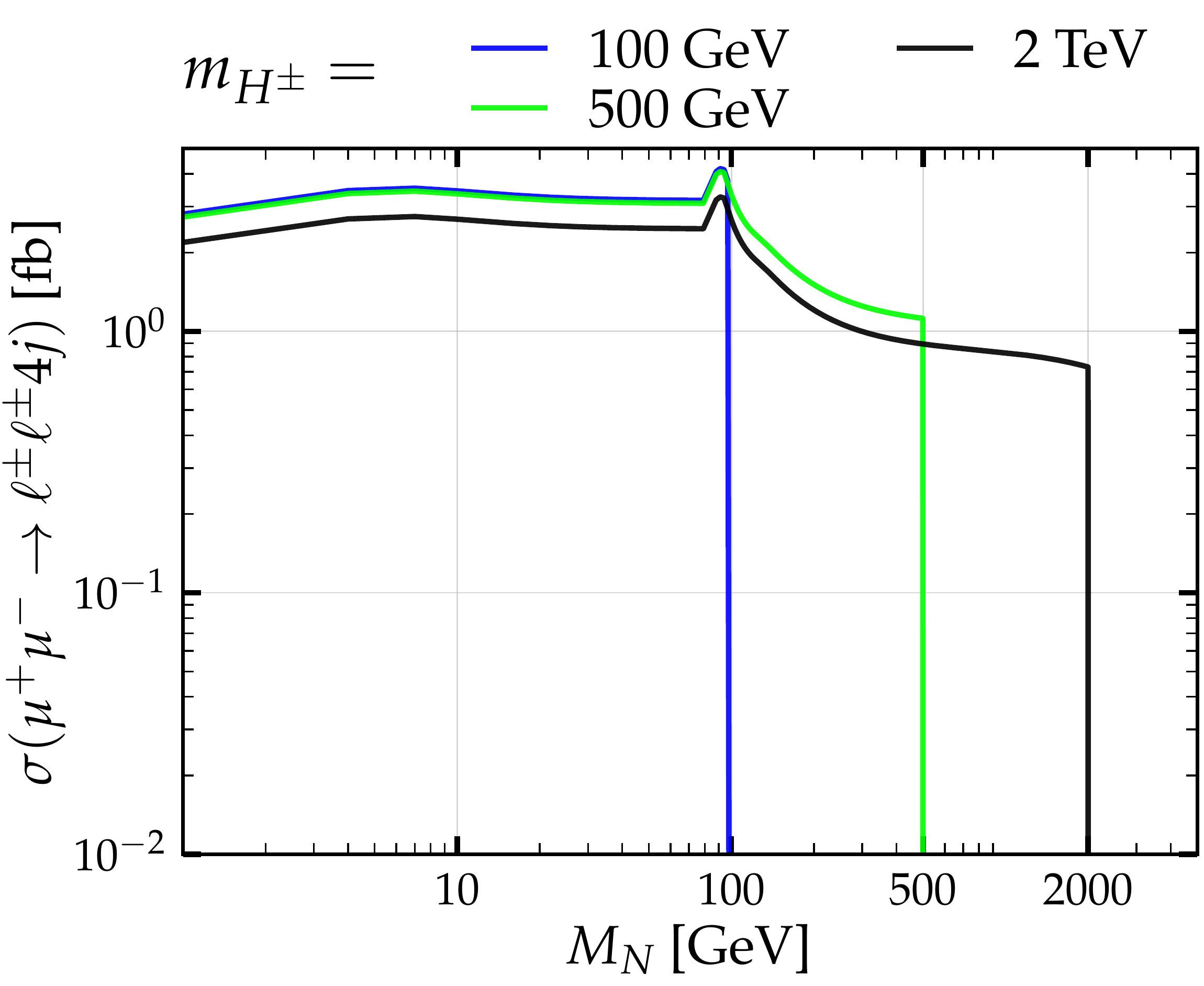}~~~	
	\includegraphics[width=0.32\linewidth]{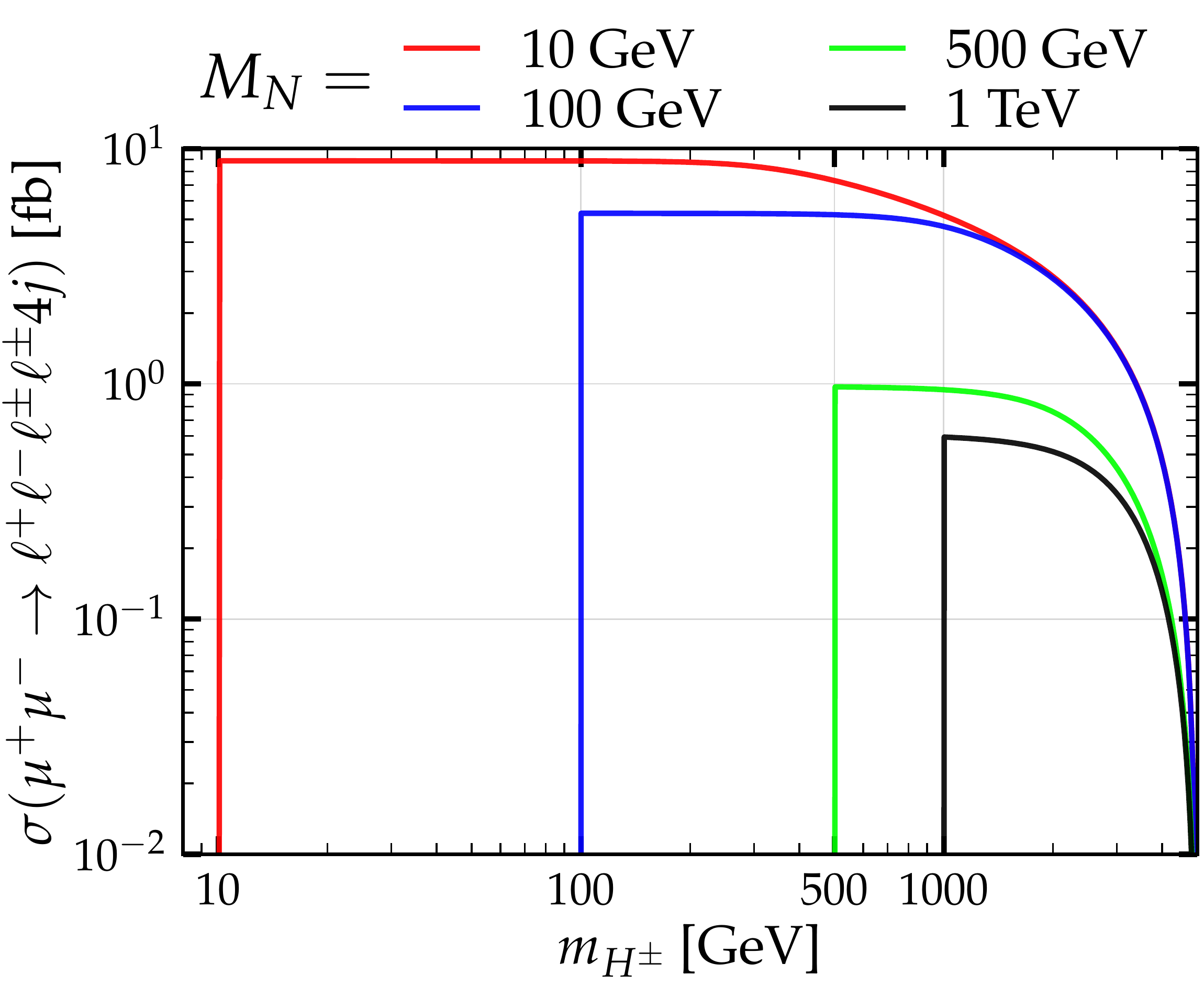}~~~
	\includegraphics[width=0.32\linewidth]{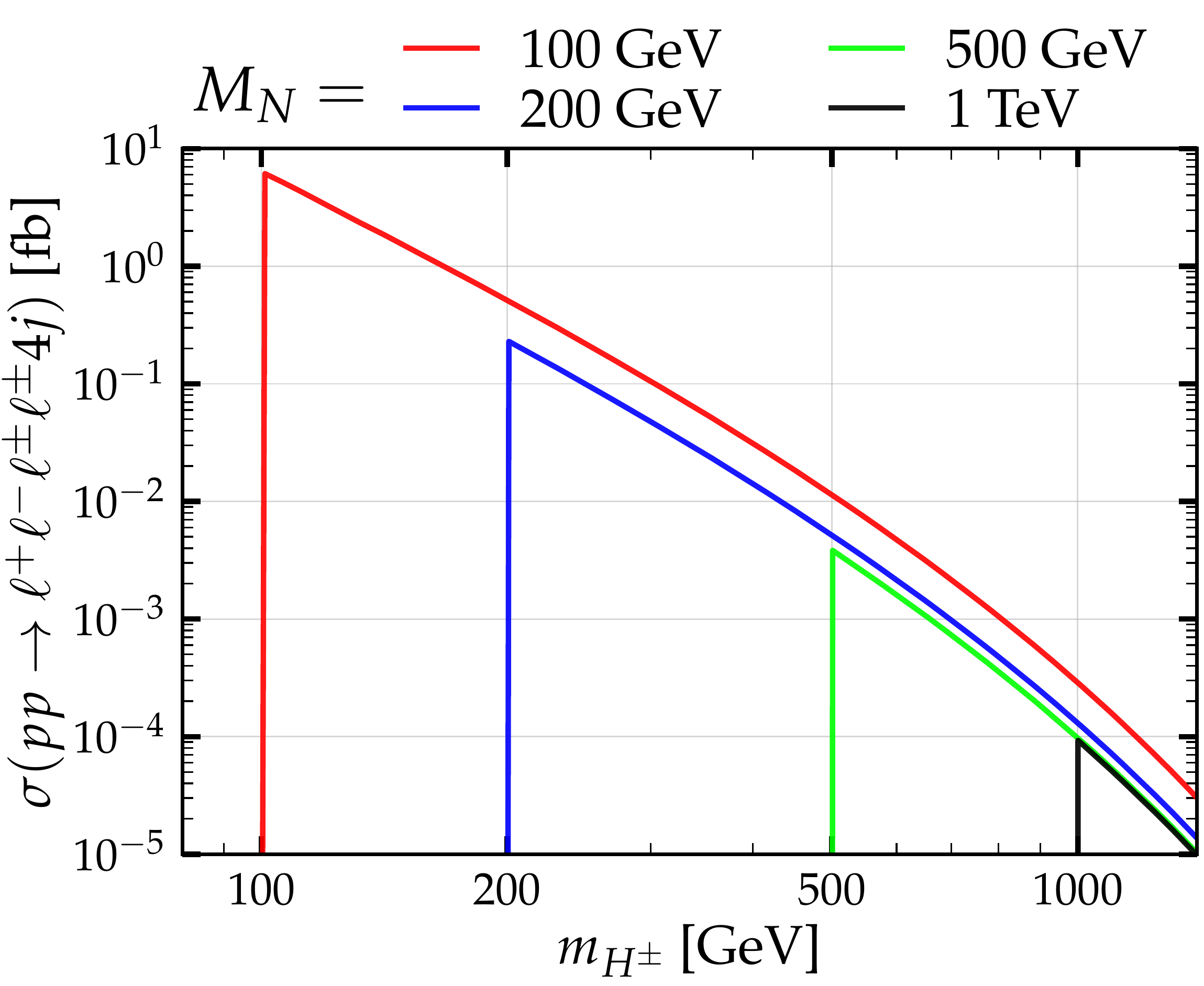}		
         \caption{
           \footnotesize{LNV cross-sections at $\sqrt{s}=$ 10 TeV for
           $\mu^+\mu^-\to N \overline{N}\to \ell^\pm \ell^\pm 4j$ versus the heavy neutrino mass $M_N$~(left)
           and for the process $\mu^+ \mu^-\to H^+H^-\to$ $\ell^+ \ell^-\ell^\pm \ell^\pm 4j$ versus the charged scalar mass~(middle).
           The right panel gives the cross-section for $pp\to H^+H^-\to$ $\ell^+\ell^-\ell^\pm \ell^\pm 4j$ versus the charged scalar mass.
           The relevant Yukawa coupling is taken as $\mathbf{Y}_S=(1,1,1)$. These rates are unsuppressed by light-heavy neutrino mixing.}}
 	\label{fig:CSmumutoee4j}
 \end{figure*}
%%
%%%%%%
 In the right panel of Fig.~\ref{fig:CSeetoee4j} we show the cross-sections for the LNV 4-lepton+4jet final states produced through the charged Higgs portal at $\sqrt{s}=3$~TeV $e^+e^-$ colider, 
assuming $|S_{\ell}|\neq 0$. As mentioned, the LNV final states can have similar cross-sections as those with LNC. 
However, in contrast to the LNV case, LNC final states suffer from SM backgrounds, requiring a dedicated analysis in order to reduce them. 
Machine learning methods such as the Boosted Decision Trees (BDT) method are often used to reduce such backgrounds~\cite{Antusch:2018bgr,Pascoli:2018heg}.
One sees that the LNV cross-sections can be large enough to deserve being explored at future lepton colliders.  
Similar results for 2-lepton+4jet and 3-lepton+4jet final states produced through $N\overline{N}$ production in $e^+ e^-$ or $NH$ production in $e^- \gamma $ collisions are shown in first two panels of Fig.~\ref{fig:CSeetoee4j}. 
%%%%%%%
 
Similar results can be obtained for the proposed muon collider~\cite{Delahaye:2019omf,Kwok:2023dck,Mekala:2023diu} for a center-of-mass energy $\sqrt{s}=10$ TeV. 
The first two panels of Fig.~\ref{fig:CSmumutoee4j} show the cross-section for LNV processes $\mu^+\mu^-\to N\overline{N}\to \ell^\pm \ell^\pm 4j$ and $\mu^+ \mu^-\to H^+H^-\to$ $\ell^+\ell^-\ell^\pm \ell^\pm 4j$, respectively. 
For a $pp$ collider, the charged Higgs can be pair-produced via the neutral current Drell-Yan mechanism, with a sizeable cross-section. 
This is seen in the last panel of Fig.~\ref{fig:CSmumutoee4j}, which shows the cross-section for the LNV process $pp\to H^+H^-\to$ $\ell^+\ell^-\ell^\pm \ell^\pm 4j$ at $\sqrt{s}=14$ TeV center-of-mass energy. 
%%%%%%%%%%
\begin{figure}[!htbp]
\includegraphics[width=0.5\linewidth]{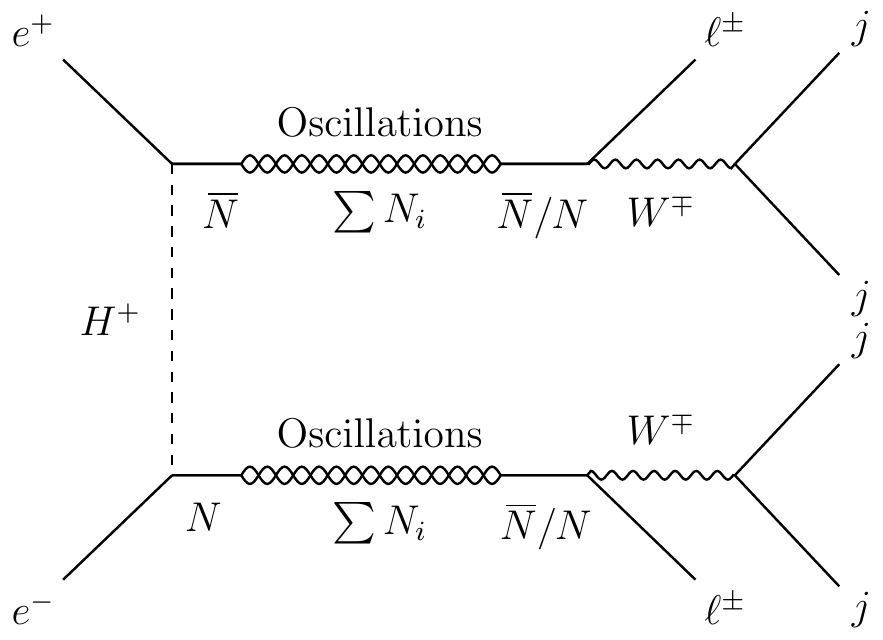}
\caption{\footnotesize{Possible LNV and LNC final states coming from the process $e^+e^-\to N\overline{N}$ via heavy neutrino-antineutrino oscillation. Because of the superposition of mass eigenstates ($\Sigma_i N_i$), $N$ oscillate into $\overline{N}$ or $N$ and decays into an antilepton ($\ell^+$) or a lepton ($\ell^-$) giving rise to a LNV or LNC signal.}}
\label{fig:N_oscillation}
\end{figure} 

%%%%%%%%%%

\section{The role of heavy neutrino-antineutrino oscillations}
\label{sec:oscillation}
\begin{figure}[!htbp]
\includegraphics[width=0.45\linewidth]{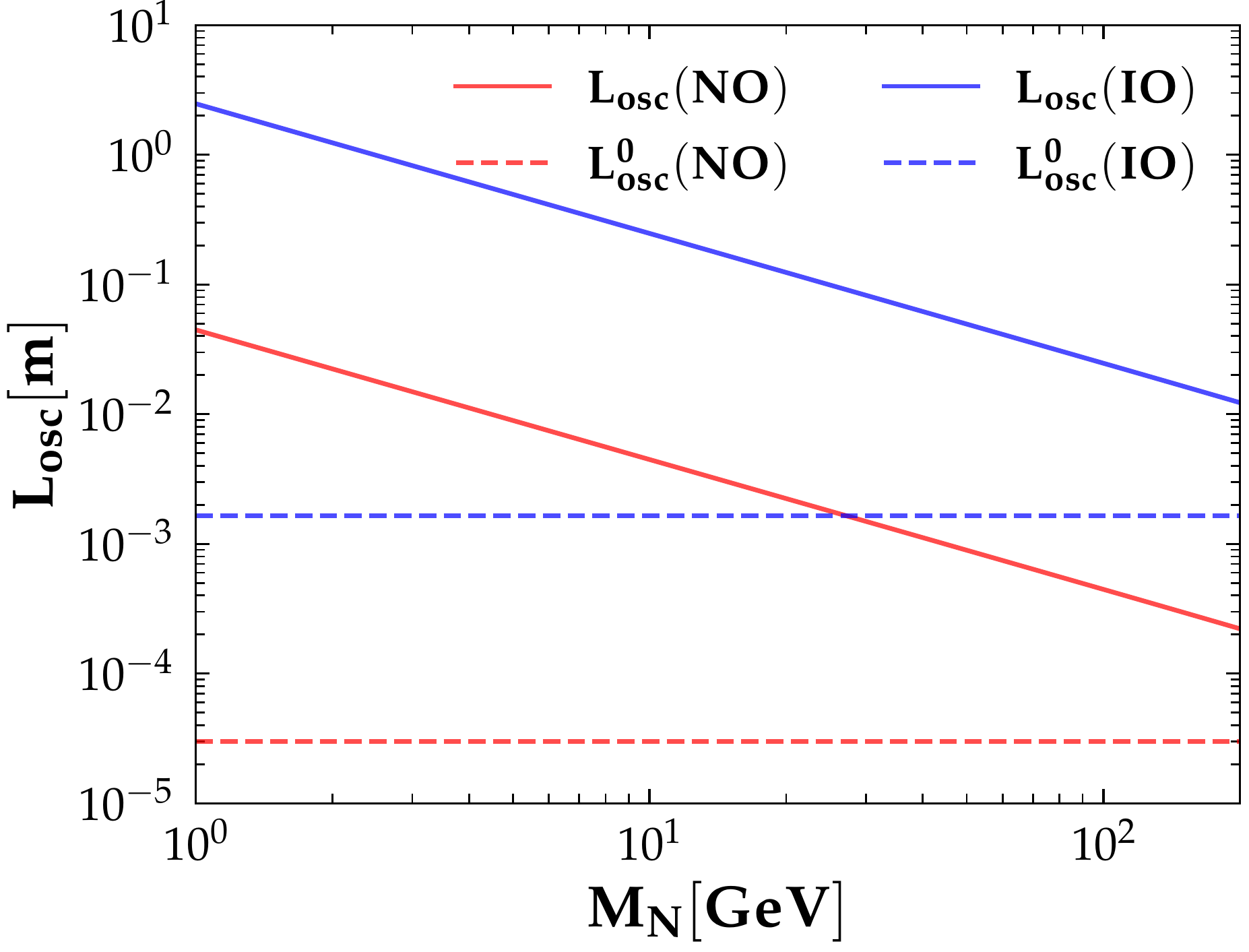}\quad
\includegraphics[width=0.45\linewidth]{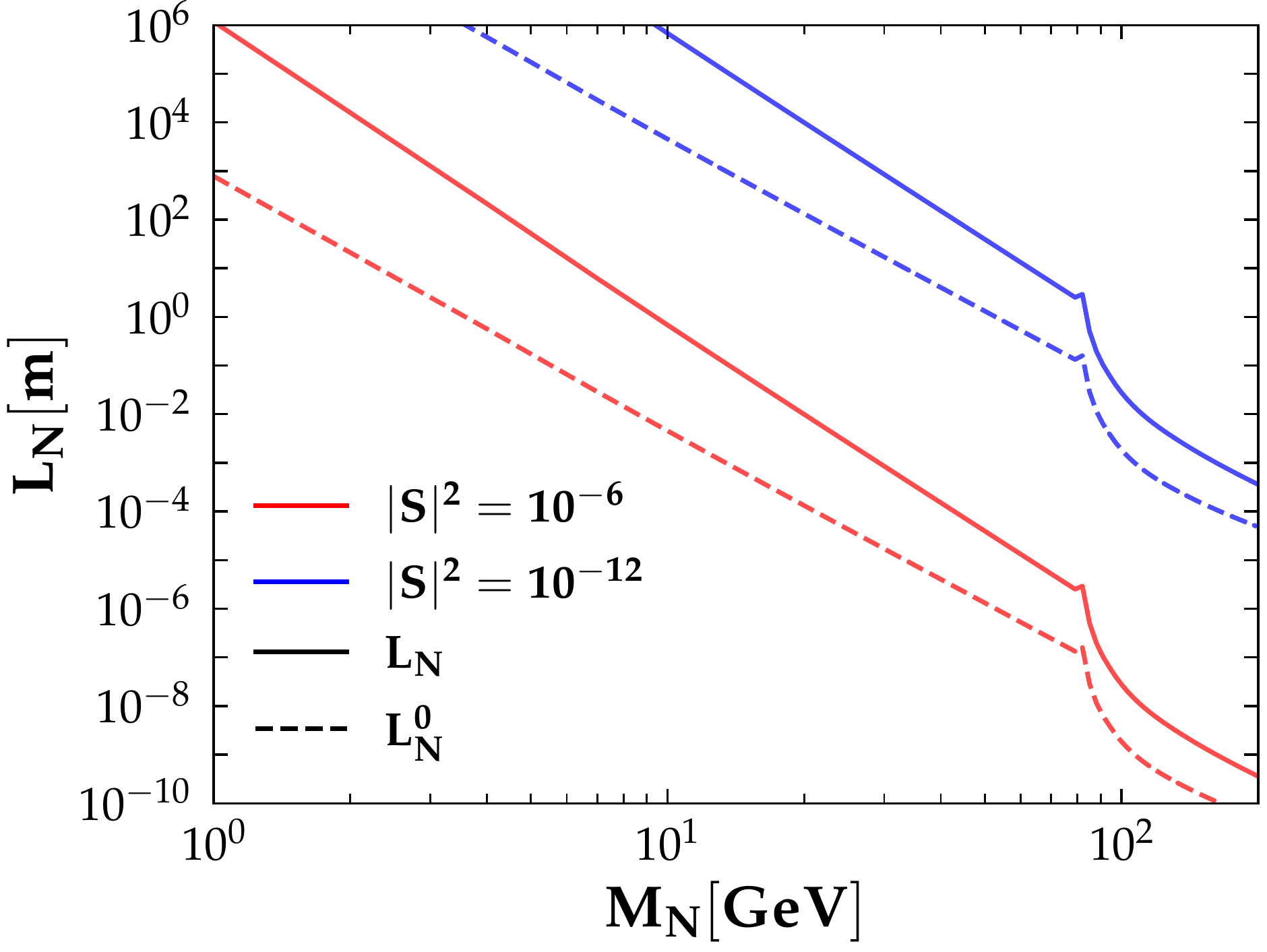}
\caption{\footnotesize{Heavy neutrino oscillation length (left panel) and the decay length (right panel) as a function of its mass $M_N$. Dashed lines refer to the proper frame, while solid lines correspond to the laboratory frame, assuming that the production process is $e^+e^-\to N\overline{N}$ with c.m. energy $\sqrt{s}=3$ TeV.}}
\label{fig:osc-length}
\end{figure}

%%%%%%%%%
Let us briefly summarise here the main features of heavy neutrino-antineutrino oscillations. The heavy neutrino-like state $N$ and anti-neutrino-like state $\bar{N}$ are written in terms of the mass eigenstates as $N(\bar{N})=(\mp i N_4 +  N_5)/\sqrt{2}$. 
Oscillations occur due to interference between the mass eigenstates $N_{4,5}$ during propagation. 
In the presence of heavy neutrino-antineutrino oscillations, one expects the following possible final states from the production process such as $e^+e^-\to N\overline{N}$,
%%%%%%%%
\begin{align}
e^+e^-\to N\overline{N} =
\begin{dcases}
N \rightsquigarrow N, \overline{N}\rightsquigarrow \overline{N}\Rightarrow \ell^+\ell^- 4j~(\text{LNC}) \\
N\rightsquigarrow \overline{N}, \overline{N}\rightsquigarrow \overline{N}\Rightarrow \ell^+\ell^+ 4j~(\text{LNV}) \\
N\rightsquigarrow N, \overline{N}\rightsquigarrow N\Rightarrow \ell^-\ell^- 4j~(\text{LNV}) \\
N\rightsquigarrow \overline{N}, \overline{N}\rightsquigarrow N\Rightarrow \ell^+\ell^- 4j~(\text{LNC}) 
\end{dcases} 
\end{align}
%%%%%%%%
 Without oscillations one has only LNC final states. However, in the presence of heavy neutrino oscillation we can have both LNV and LNC final states, see Fig.~\ref{fig:N_oscillation}. 
 The oscillation probabilities of $N(0)\to N(\tau)/\bar{N}(\tau)$ as a function of the proper time $\tau$ are given as
%%%%%%
\begin{align}
\bar{P}_{\rm osc}^{N\to N(\bar{N})}(\tau)=|\braket{N(\bar{N})|N(\tau)}|^2=|g_{\pm}(\tau)|^2, \text{  with  } |g_{\pm}(\tau)|^2=\frac{e^{-\Gamma_N\tau}}{2} \left(1\pm \cos (\Delta M\tau)\right).
\label{eq:Noscillation}
\end{align}
%%%%%%
The oscillation period is $\tau_{\rm osc}=2\pi/\Delta M$ and the oscillation length is given by $L_{\rm osc}^0=c\tau_{\rm osc}$. The oscillation expression in Eq.~\ref{eq:Noscillation} takes the following form in the laboratory frame,
%%%%%%
\begin{align}
\bar{P}_{\rm osc}^{N\to N(\bar{N})}(x) =\frac{1}{2} e^{-x/L_N}\left(1\pm \cos (2\pi x/L_{\rm osc})\right),
\end{align}
%%%%%%
where $L_N=L_N^0\sqrt{\gamma^2-1}$ and $L_{\rm osc}=L_{\rm osc}^0\sqrt{\gamma^2-1}$ are the heavy neutrino decay length and oscillation length in the laboratory frame, where $\gamma=E_N/M_N$ is the Lorentz factor for the specific process. 
In our minimal linear seesaw mechanism, the mass-splitting is $\Delta M=41.51\times 10^{-3}$ eV for normal mass ordering~(\textbf{NO}), and $\Delta M=749.8\times 10^{-6}$ eV for inverted ordering~(\textbf{IO}). 
The corresponding oscillation lengths~($L_{\rm osc}^0$) are listed in Table.~\ref{tab:benchmark-models}. If the Lorentz factor is large, the oscillation length in the laboratory frame can be large enough to be experimentally resolvable~\cite{Antusch:2017ebe}. 
In Table.~\ref{tab:benchmark-models}, we show the oscillation length in the laboratory frame for a heavy neutrino mass $M_N=10$ GeV~(4th column) and $M_N=50$ GeV~(5th column), assuming the production process is $e^+e^-\to N\overline{N}$ with c.m. energy $\sqrt{s}=3$ TeV.
In Fig.~\ref{fig:osc-length}, left panel, we show the oscillation length in the laboratory frame along with the proper frame as a function of the heavy neutrino mass $M_N$. We see that \textbf{IO} looks particularly promising in this context. In order to observe this oscillation pattern, it is also important that decay of the heavy neutrinos is sufficiently displaced from the primary vertex. This can be easily achieved as the decay length covers a wide range of possible values, depending on the choices of $M_N$ and $|S|^2$, as shown in the right panel of Fig.~\ref{fig:osc-length}.

\begin{table}
\begin{tabular}{|c|c|c|c|c|}
\hline
\textbf{}              & $\Delta M \,[\text{eV}]$           & $L_\text{osc}^0 \,[\text{m}] $ &  $L_\text{osc} \,[\text{m}]\,(\gamma=150) $ & $L_\text{osc} \,[\text{m}]\,(\gamma=30)$ \\
\hline
\textbf{NO}              & $41.51\times 10^{-3}$                                & $2.98\times 10^{-5}$ &  $4.5\times 10^{-3} $  & $8.9\times 10^{-4} $  \\
\hline
\textbf{IO}              & $749.8\times 10^{-6}$                                & $1.65\times 10^{-3}$ &   $247\times 10^{-3}$  &  $49.5\times 10^{-3} $ \\
\hline
\end{tabular}
\caption{\footnotesize{Expected heavy neutrino oscillation length in the proper frame~($L_{\rm osc}^0$) and laboratory frame~($L_{\rm osc}$), for both the \textbf{NO} and \textbf{IO} cases. The laboratory frame oscillation lengths assume $e^+e^-\to N\overline{N}$ for masses $M_N=10$~GeV~($\gamma=150$) and $M_N=50$~GeV~($\gamma=30$) and c.m. energy $\sqrt{s}=3$ TeV.}}
\label{tab:benchmark-models}
\end{table}
%%%%%%
%%%%%%
\begin{figure}[!htbp]
\includegraphics[width=0.46\linewidth]{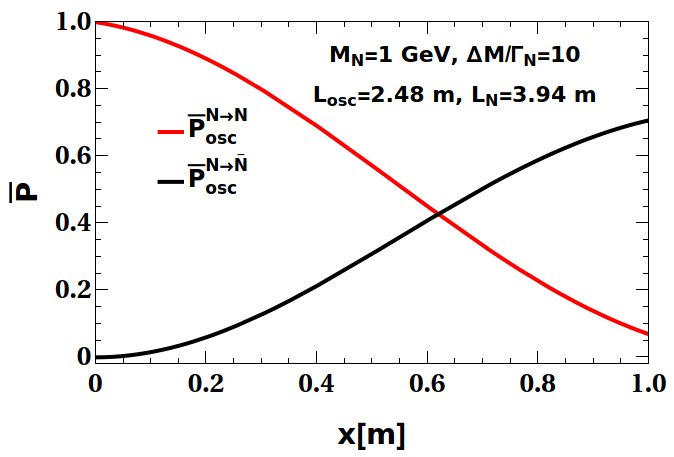}~
\includegraphics[width=0.46\linewidth]{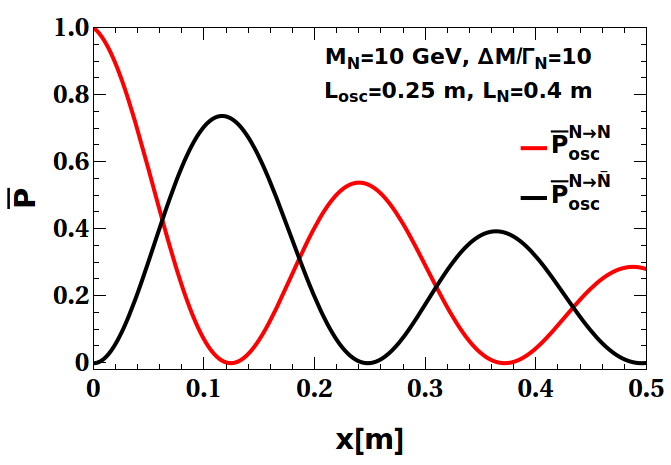}
\includegraphics[width=0.46\linewidth]{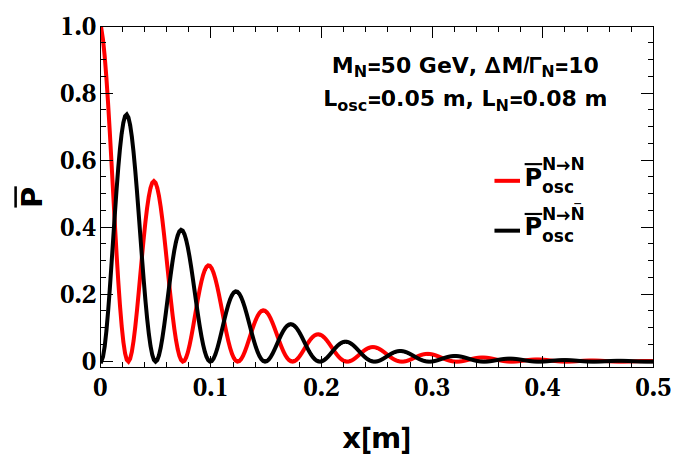}~
\includegraphics[width=0.46\linewidth]{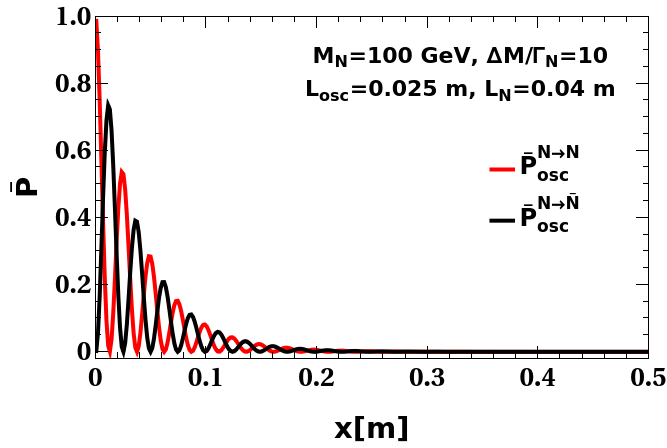}
\caption{\footnotesize{The oscillation probabilities $\bar{P}_{\rm osc}^{N\to N}(x)$~(red line) and $\bar{P}_{\rm osc}^{N\to \overline{N}}(x)$~(black line) as a function of distance in the laboratory frame, assuming $e^+e^-\to N\overline{N}$ with c.m. energy $\sqrt{s}=3$ TeV as the production process. 
The upper left, upper right, lower left, and lower right panels correspond to $M_N=1$ GeV, 10 GeV, 50 GeV, and 100 GeV, respectively, with \textbf{IO}, mass-splitting $\Delta M=749.8\times 10^{-6}$ eV and $\Delta M/\Gamma_N=10$.}}
\label{fig:prob}
\end{figure} 
%%%%%
In Fig.~\ref{fig:prob}, we show the oscillation probabilities $\bar{P}_{\rm osc}^{N\to N}(x)$~(red line) and $\bar{P}_{\rm osc}^{N\to \overline{N}}(x)$~(black line) as a function of distance in the laboratory frame for \textbf{IO} with mass-splitting $\Delta M=749.8\times 10^{-6}$ eV and $\Delta M/\Gamma_N=10$. The four panels correspond to $M_N=1$ GeV, 10 GeV, 50 GeV and 100 GeV, respectively. 
For each panel in Fig.~\ref{fig:prob}, the oscillation length in the rest frame is $L_{\rm osc}^0=1.65\times 10^{-3}$ m, but in the laboratory frame it changes depending on the heavy neutrino mass $M_N$. Smaller $M_N$ values have larger boost factor and accordingly also larger oscillation length. 
This is why the oscillation period is relatively larger for $M_N=1$ GeV and 10 GeV, but becomes short for $M_N=50$ GeV and $M_N=100$ GeV. We note that the oscillation length in the {\bf NO} case is shorter compared to {\bf IO}. This maybe  still resolvable with the FCC-ee or ILC
tracking resolution, though more challenging than for the {\bf IO} case.

The total probability for an unstable and oscillating particle $N$ to decay in a proper time window is given by
%%%%%%%
\begin{align}
 P_{\rm osc}^{N\to N(\bar{N})}(\tau_{\rm min},\tau_{\rm max})=\Gamma_N\int_{\tau_{\rm min}}^{\tau_{\rm max}} \bar{P}_{\rm osc}^{N\to N(\bar{N})}(\tau) d\tau,
\end{align}
%%%%%%
which takes the following form in the laboratory frame,
%%%%%
\begin{align}
 P_{\rm osc}^{N\to N(\bar{N})}(x_1,x_2)=\frac{1}{L_N}\int_{x_1}^{x_2} \bar{P}_{\rm osc}^{N\to N(\bar{N})}(x) dx.
\end{align}
%%%%%%%
The expected number of heavy neutrinos produced at the interaction point and decaying with a displacement of at least $x_1$ and at most $x_2$, can be obtained by combining the production and decay of heavy neutrinos~\footnote{The vertex displacement $x$ is defined as the distance between the primary vertex where the heavy neutrino~($N$ or $\bar{N}$) with finite lifetime was produced, and its secondary decay vertex. %%where the produced heavy neutrinos decay into a number of daughter particles.
}. 
Hence, the expected number of LNC and LNV events is,
%%%%%
\begin{align}
& N^{\rm LNC}(x_1,x_2,\sqrt{s},\mathcal{L})=\mathcal{L}\, \sigma\, \text{BR} \, \Big[\big(P_{\rm osc}^{N\to N}(x_1,x_2)\big)^2 + \big(P_{\rm osc}^{N\to \overline{N}}(x_1,x_2)\big)^2 \Big], \\
& N^{\rm LNV}(x_1,x_2,\sqrt{s},\mathcal{L})=2\mathcal{L}\, \sigma\, \text{BR} \, P_{\rm osc}^{N\to N}(x_1,x_2) P_{\rm osc}^{N\to \overline{N}}(x_1,x_2),
\end{align}
%%%%%%%
where $\mathcal{L}$ is the luminosity of the collider, $\sigma$ and BR are the heavy neutrino production cross-section and relevant decay branching ratio~\footnote{For the case of a realistic  collider experiment we can estimate the expected event number as $N^{\rm LNV}=\mathcal{L}\, \sigma\, \text{BR}\,\int D(\theta,\gamma)\, P_{\rm osc}^{N\to N}\left(x_1(\theta),x_2(\theta)\right) P_{\rm osc}^{N\to \overline{N}}\left(x_1(\theta),x_2(\theta)\right)\,d\theta\,d\gamma$, where $D(\theta,\gamma)$ is the probability that heavy neutrinos are produced with the boost $\gamma$ and with angle $\theta$ with respect to the beam axis.}. \par 

In deriving the above expressions, we use the following relations, $P_{\rm osc}^{\overline{N}\to \overline{N}}=P_{\rm osc}^{N\to N}$ and $P_{\rm osc}^{\overline{N}\to N}=P_{\rm osc}^{N\to \overline{N}}$. 
Note that, unlike all other type-I seesaw schemes, the primordial heavy neutrino production rates here are not controlled by the (neutrino-mass suppressed) light-heavy mixing. In contrast, it is proportional to the Yukawa coupling $Y_S$ which is not suppressed by the neutrino mass restrictions.
%%%%%%%

Note that, even if the parameters do not allow for the heavy-neutrino oscillations to be resolvable, an integrated effect could still be measured. In the following we discuss this integrated effect on the expected number of LNV and LNC events for an idealized infinite detector, as well as a realistic finite one.

%%%%%%
\subsection{Infinite detector}
\label{subsec:infinite-detector}
%%%%%%
If the detector size is infinite, the experiment can observe all decays from the origin to infinity, so one can take $\tau_{\rm min}\to 0$ and $\tau_{\rm max}\to\infty$, and the oscillation probability takes the following form,
%%%%%%
\begin{align}
 P_{\rm osc}^{N\to N}(0,\infty) = \frac{1}{2}\left(
\frac{\Gamma_N^2}{\Delta M^2 + \Gamma_N^2} + 1\right), \text{  and  }
P_{\rm osc}^{N\to \bar{N}}(0,\infty) = \frac{1}{2}
\frac{\Delta M^2}{\Delta M^2 + \Gamma_N^2}.
\end{align}
%%%%%%%%
Therefore, the ratio between the LNC and LNV decay modes when the heavy quasi-Dirac neutrinos are produced as real particles is given by~\cite{Anamiati:2016uxp,Antusch:2022ceb}
%%%%%%%%
\begin{align}
R_{\ell\ell}=\frac{N^{\rm LNV}}{N^{\rm LNC}}=\frac{y^2(2+y^2)}{2+y^2(2+y^2)} \text{  with  } y=\frac{\Delta M}{\Gamma_N}.
\end{align}
%%%%%%%%%%
\begin{figure}[!htbp]
\includegraphics[width=0.46\linewidth]{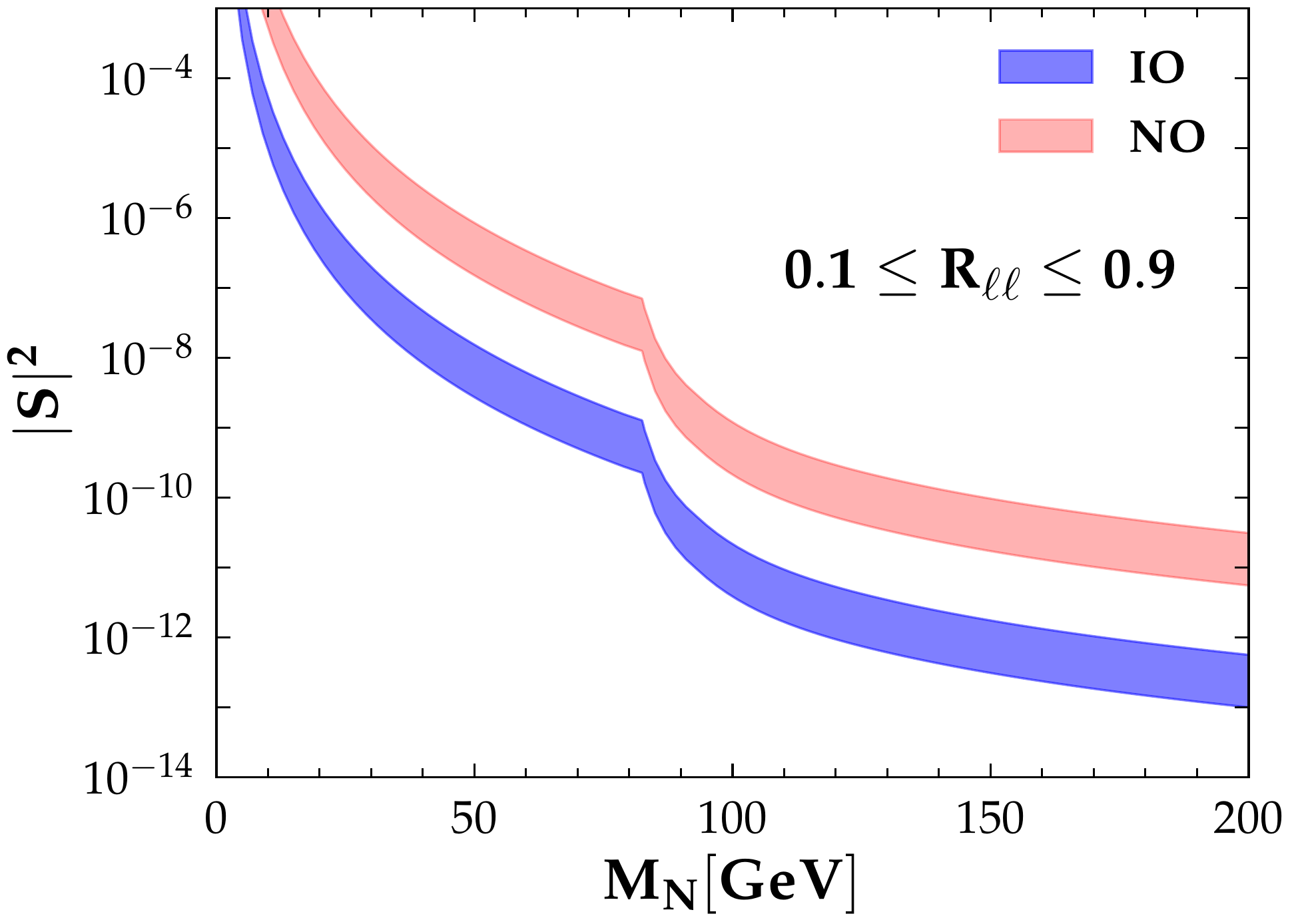}~~~~~
\includegraphics[width=0.425\linewidth]{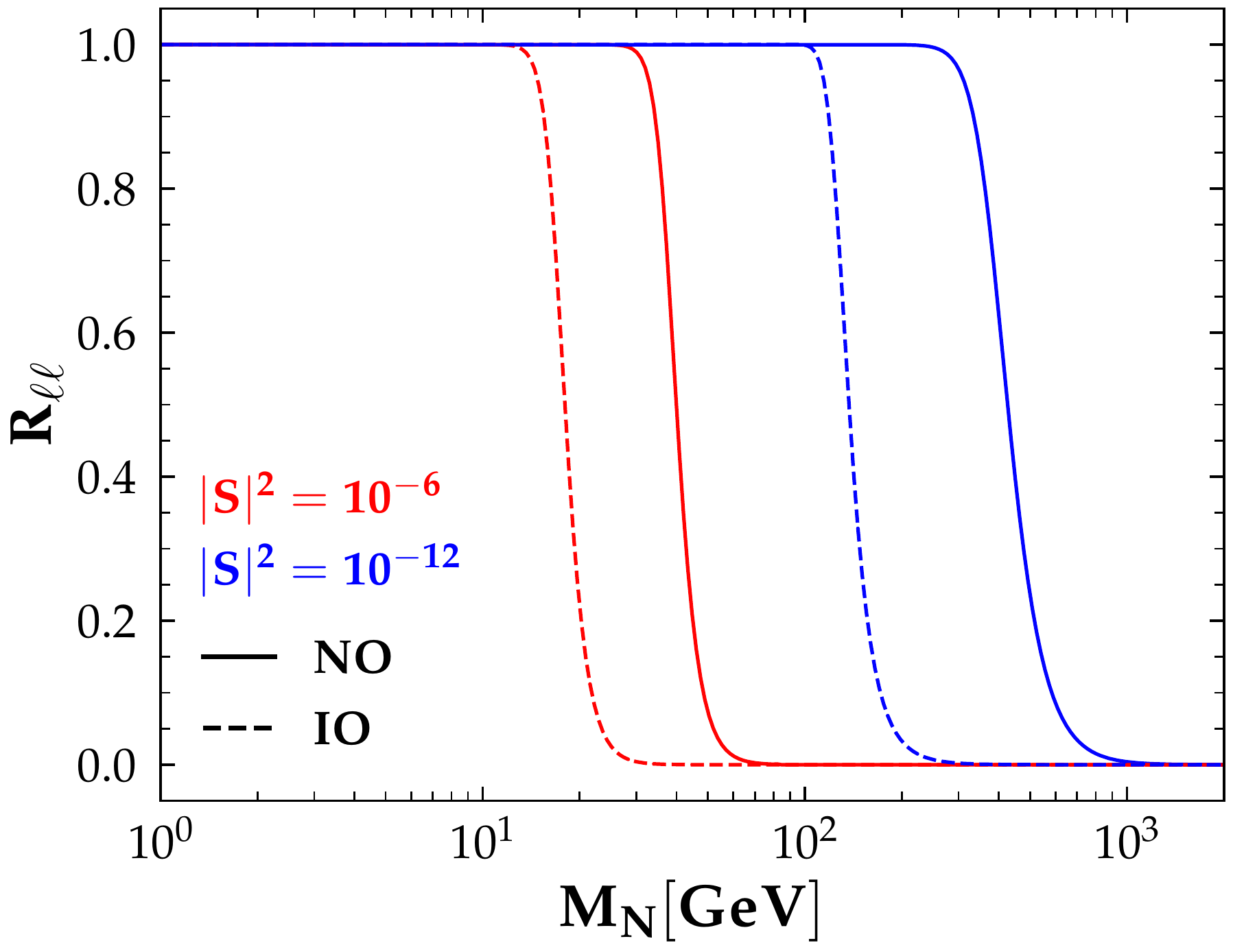}
\caption{\footnotesize{Left Panel: The red and blue bands give the regions in $|{\bf S}|^2-M_N$ leading to $0.1\leq R_{\ell\ell}\leq 0.9$  for {\bf NO} and {\bf IO}, respectively. Right Panel: $R_{\ell\ell}$ vs $M_N$ for mixing parameter values $|S|^2= 10^{-6}$ and $10^{-12}$.}}
\label{fig:Rll}
\end{figure} 
%%%%%%%
\begin{figure}[!htbp]
\includegraphics[width=0.99\linewidth]{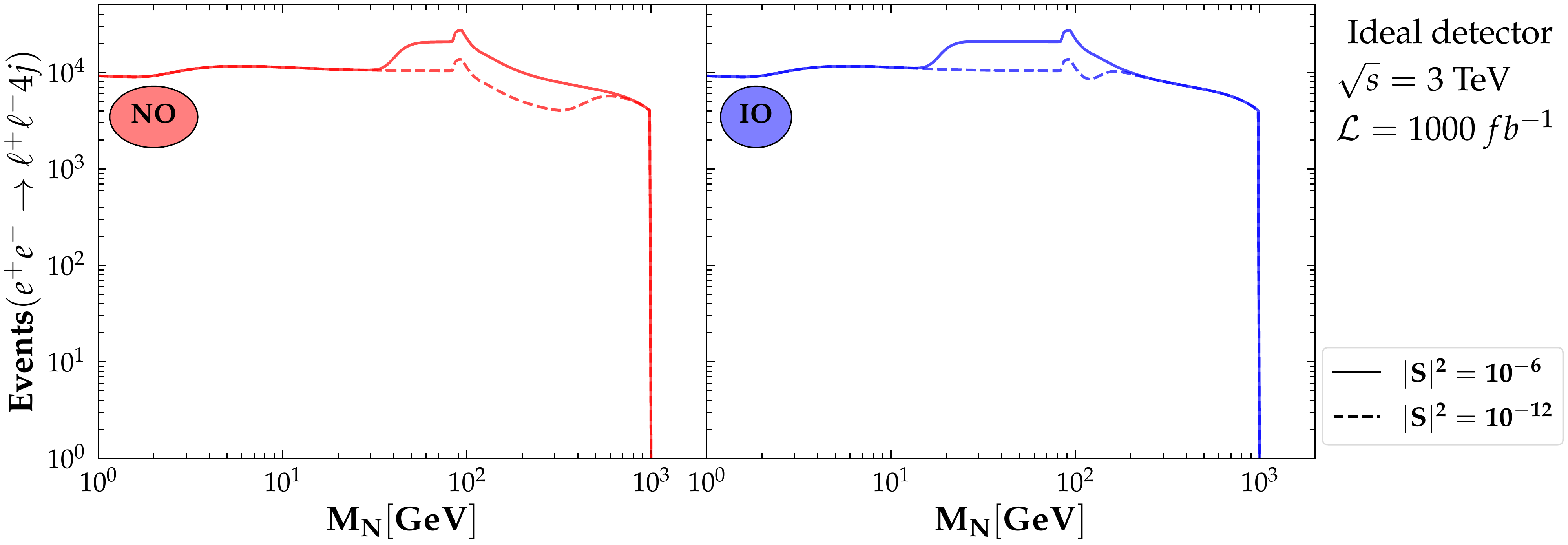}
\includegraphics[width=0.99\linewidth]{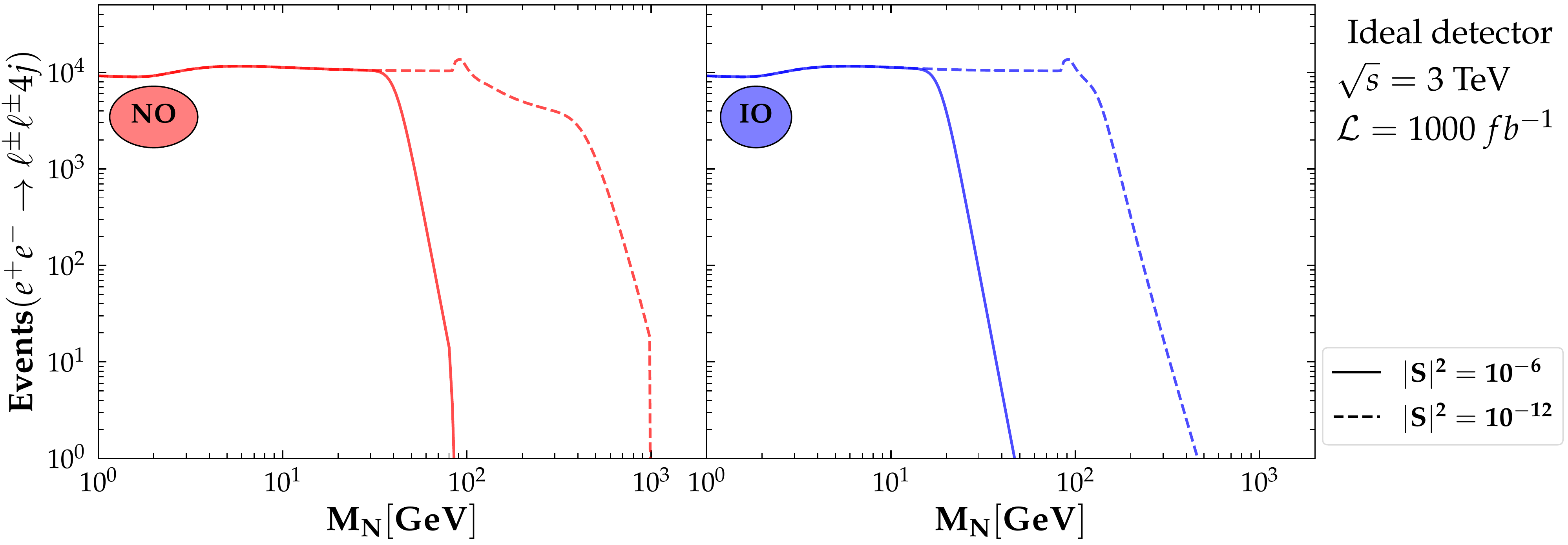}
\caption{\footnotesize{Expected number of LNC~($e^+ e^-\to \ell^\pm \ell^\mp 4j$) (upper panel) and LNV~($e^+ e^-\to \ell^\pm \ell^\pm 4j$) (lower panel) events assuming an idealized lepton collider with c.m. energy $\sqrt{s}=3$ TeV for fixed mixing $|{\bf S}|^2=10^{-6}$~(solid) and $|{\bf S}|^2=10^{-12}$~(dashed). Mass of the charged Higgs ($H^\pm$) is taken to be 1 TeV and $\ell=e,\mu$ and $\tau$. 
%When $R_{\ell\ell}\to0$, the number of LNC events become twice the number of LNC/LNV events when $R_{\ell\ell}=1$ as the total number of events is same for the Dirac and Majorana cases~\cite{Drewes:2022rsk}.
}}
\label{fig:ideal-events}
\end{figure} 

Hence one finds that $R_{\ell\ell}=1$ for the case of standard type-I high-scale seesaw, while for low-scale seesaw schemes $R_{\ell\ell}$ lies anywhere within the $[0:1]$ range, depending on the $\Delta M/\Gamma_N$ ratio. %%\par
When the mass splitting $\Delta M$ is larger than a few times the width $\Gamma_N$~($y\gtrsim 4$), one finds that $R_{\ell\ell}$ approaches rapidly the limit $R_{\ell\ell}=1$. 
For example, the left panel of Fig.~\ref{fig:Rll} shows the value of the mixing factor $|{\bf S}|^2$ and mass $M_N$ required to achieve an LNV to LNC ratio in the range $0.1 \leq R_{\ell\ell}\leq 0.9$ for both {\bf NO}~(red) and {\bf IO}~(blue). 
In the right panel of Fig.~\ref{fig:Rll} we show the behaviour of $R_{\ell\ell}$ as a function of $M_N$ for the light-heavy mixing parameter ${\bf|S|^2}=10^{-6}$ (red line) and $10^{-12}$ (blue line). The solid and dashed lines represent the {\bf NO} and {\bf IO} cases, respectively. 

Note that the smaller the mixing $|{\bf S}|^2$, the larger the range of $M_N$ up to which the condition $R_{\ell\ell}\approx 1$ can be satisfied. 
This is due to the fact that the $\Gamma_N\propto |{\bf S}|^2$, hence the ratio $\Delta M/\Gamma_N$ can be sizeable even for larger values of $M_N$. 
One also sees that for the {\bf NO}, the condition of $R_{\ell\ell}\approx 1$ can be realized for larger range of $M_N$ compared to the case of {\bf IO}. This is due to the fact that $\Delta M^{\rm {\bf NO}}>\Delta M^{\rm {\bf IO}}$.
The upshot is that within our linear seesaw setup there is a wide parameter region with $\Delta M \geq \Gamma_N$, so that the ratio $R_{\ell\ell}$ can have a non-negligible value.

In Fig.~\ref{fig:ideal-events}, we show the expected number of LNC~($e^+ e^-\to \ell^\pm \ell^\mp 4j$) (upper panel) and LNV~($e^+ e^-\to \ell^\pm \ell^\pm 4j$) (lower panel) events at lepton colliders with c.m. energy $\sqrt{s}=3$~TeV as a function of heavy neutrino mass $M_N$ for the {\bf NO}~(left panel) and {\bf IO}~(right panel). The solid and dashed lines correspond to different values of the light-heavy mixing parameter $|{\bf S}|^2=10^{-6}$ and $|{\bf S}|^2=10^{-12}$, respectively. We fix the luminosity to be $\mathcal{L}=1000\,\text{fb}^{-1}$ and the Yukawa coupling $|{\bf Y_S}|=1$.
One sees that for an ideal detector~($x_1=0$ and $x_2=\infty$) one can have a large number of LNV signal events, as long as $R_{\ell\ell}$ is non-negligible. As we discussed earlier, for fixed light-heavy mixing, above a certain value of the heavy neutrino mass $M_N$, the ratio $\Delta M/\Gamma_N$ will become small and, as a result, $R_{\ell\ell}$ will also become smaller. 
Hence, unlike the LNC events, the LNV event numbers drop above certain $M_N$ value, for fixed mixing $|{\bf S}|$. For {\bf IO}, this drop in LNV event numbers happens earlier compared to {\bf NO} case due to the fact that $\Delta M^{{\bf NO}} > \Delta M^{{\bf IO}}$ and hence $R_{\ell\ell}^{{\bf NO}}>R_{\ell\ell}^{{\bf IO}}$ for fixed mixing $|{\bf S}|$ and mass $M_N$. 

\subsection{Finite detector}
\label{subsec:finite-detector}
%%%%%%%%
The above picture is significantly altered when the finite size of the experimental setup is taken into consideration, by integrating only over the fiducial detector size,
%%%%%%%%
\begin{align}
R_{\ell\ell}^{\rm obs}(x_1,x_2)=\frac{N^{\rm LNV}}{N^{\rm LNC}}=\frac{2 P_{\rm osc}^{N\to N}(x_1,x_2) P_{\rm osc}^{N\to \overline{N}}(x_1,x_2)}{\big(P_{\rm osc}^{N\to N}(x_1,x_2)\big)^2 + \big(P_{\rm osc}^{N\to \overline{N}}(x_1,x_2)\big)^2 }.
\end{align}
%%%%%%
At lepton colliders such as FCC-ee or CEPC we can reliably estimate the total number of heavy neutrino decays inside a cylindrical detector of length $l_{\rm cyl}$ and diameter $d_{\rm cyl}$ by  setting $x_2=(1/2)\, (3/2)^{1/3} d_{\rm cyl}^{2/3}l_{\rm cyl}^{1/3}$, so that a sphere of radius $x_2$ has the same volume as the cylinder~\cite{Drewes:2022rsk}. Following Ref.~\cite{Drewes:2022rsk}, we find $x_2\approx \mathcal{O}(1\,\text{m})$ and accordingly we take here $x_2=1\,\text{m}$ to be definitive.
%%%%%%%
\begin{figure}[!htbp]
\includegraphics[width=0.99\linewidth]{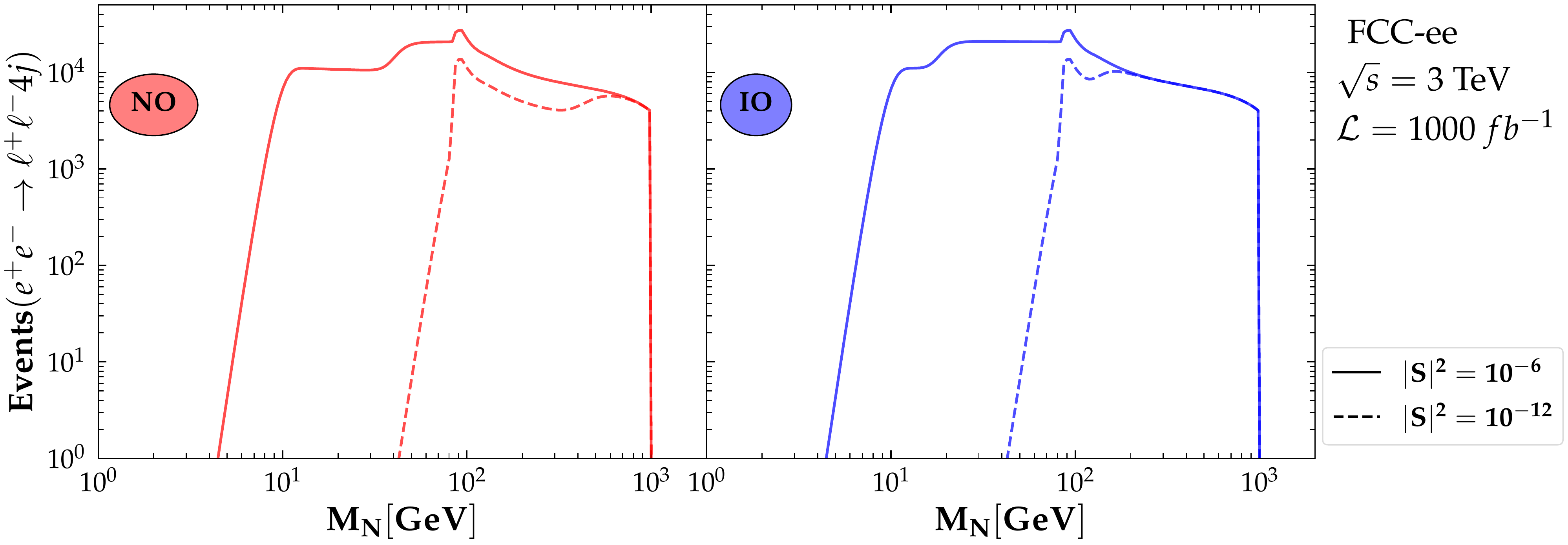}
\includegraphics[width=0.99\linewidth]{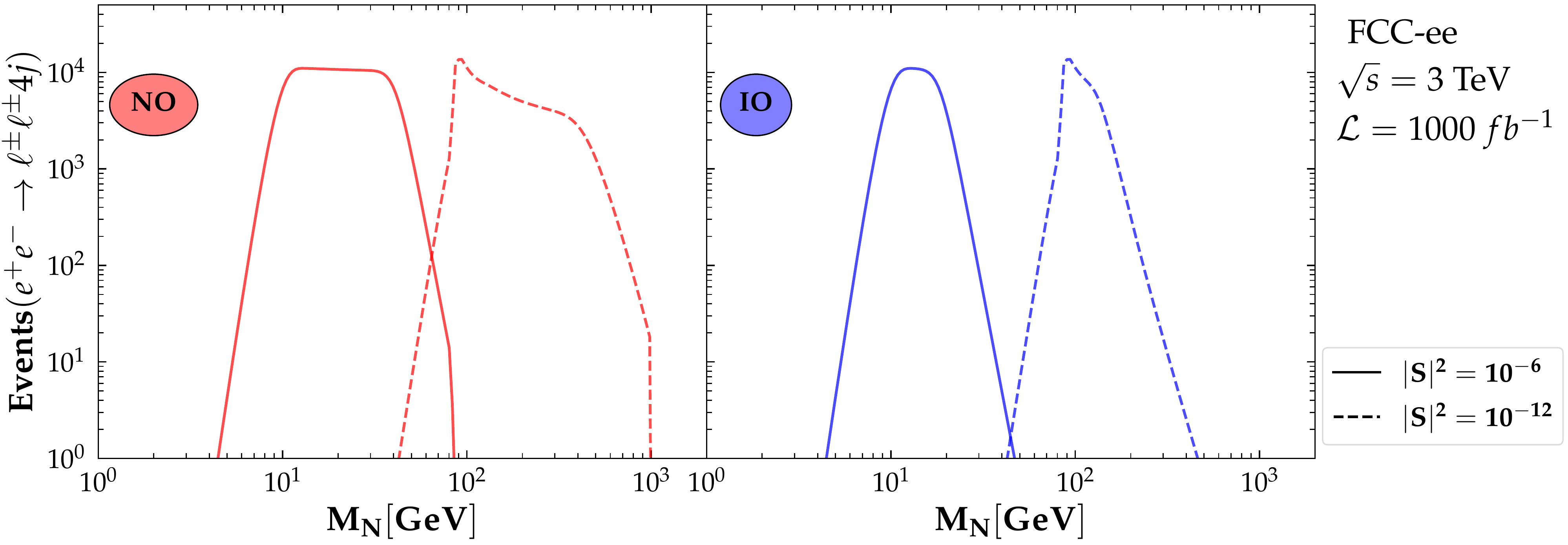}
\caption{\footnotesize{Similar to Fig.~\ref{fig:ideal-events} but now for a FCC-ee-like detector. We take the detector length to be $\mathcal{O}(1\,\text{m})$.}}
\label{fig:Nevent}
\end{figure} 

In Fig.~\ref{fig:Nevent}, we show how the results change compared to Fig.~\ref{fig:ideal-events} in the case of a finite detector size. As in the case of ideal detector, the LNV events drop above a certain value of $M_N$, as the heavy neutrino oscillation is no longer effective~($\Delta M<\Gamma_N$). 
However, in addition to this drop in LNV events above a certain value of $M_N$, we have now a region where both the LNV and LNC event numbers drop. This happens below a certain value of $M_N$, due to the fact that for fixed mixing, below certain value of $M_N$, the heavy neutrino lifetime is so large that the decay length exceeds the detector length and hence heavy neutrinos will mostly decay outside the detector. %

In view of the above discussion, the following expression for LNV/LNC event numbers holds for generic seesaw models,
%%%%%%%
\begin{align}
N^{\rm LNV}=N^{\rm LNC}\, R_{\ell\ell}^{\rm obs},
\end{align}
%%%%%%%
where $R_{\ell\ell}^{\rm obs}$ takes into account the efficiency of heavy neutrino-antineutrino oscillation in a particular experiment. %%
For conventional high-scale seesaw models with heavy Majorana neutrinos $R_{\ell\ell}^{\rm obs}=1$ and hence $N^{\rm LNV}=N^{\rm LNC}$, whereas for quasi-Dirac heavy neutrinos $R_{\ell\ell}^{\rm obs}$ lies anywhere in the range $[0:1]$ depending on the detector geometry and $\Delta M/\Gamma_N$ ratio. 
Heavy neutrino production is neutrino-mass-suppressed, as it involves light-heavy neutrino mixing, %proportional to $|{\bf S}|^{2n}$ where $n\geq 1$. Hence, we 
thus one expects very few LNC events and even lesser number of LNV events if $R_{\ell\ell}^{\rm obs}\ll 1$. 
The key advantage of our linear seesaw scheme with respect to other low-scale scenarios~(see for example Refs.~\cite{Antusch:2022ceb,Antusch:2023nqd}), is that the heavy neutrino production is not neutrino-mass-suppressed, as it is determined by the $\mathbf{Y}_S$ Yukawa coupling instead of the small light-heavy neutrino mixing, allowing for sizeable LNV rates even for $R_{\ell\ell}^{\rm obs}\ll 1$.

%%%%%
\section{Lepton number violation at low energies} 
\label{sec:ovbb}
As we saw, the linear seesaw setup harbors potentially detectable LNV processes at high energies.
Can they be used to infer the Majorana nature of neutrinos, similar to neutrinoless double beta decay~($0\nu\beta\beta$)~\cite{GERDA:2019ivs}? 
In this case we have the black-box theorem~\cite{Schechter:1981bd}, which states that the observation of \znbb decay would imply that at least one neutrino is Majorana-type. 
In a low-scale seesaw the \znbb amplitude is proportional to 
\begin{align}
\braket{m_{\beta\beta}}\propto \Big|\sum_{i=1}^3 U_{ei}^2 m_{\nu_i} + \braket{p^2}\sum_{i=1}^2 \frac{K_{ei}^2}{M_{N_{i+3}}}\Big|^2,
\label{eq:dbeta-decay}
\end{align}
where the first term is the usual long-range contribution coming from light neutrino exchange, while the second is the short-range term arising from heavy neutrino exchange.
Here $U$ denotes the lepton mixing matrix determined in oscillation experiments.  
The heavy-neutrino exchange term involves the matrix block $K_{e}=(-\frac{i}{\sqrt{2}} S_e, \frac{1}{\sqrt{2}} S_e)$ characterizing the light-heavy admixtures,
and $\braket{p^2}$ is the average squared momentum-exchange. One finds that the last contribution involves
\begin{align}
\sum_{i=1}^2 \frac{K_{ei}^2}{M_{N_{i+3}}}=\frac{K_{e1}^2}{M_{N_4}}+\frac{K_{e2}^2}{M_{N_5}}= -\frac{S_{e}^2}{2} \frac{\Delta M}{M_{N_4} M_{N_5}},
\end{align}
where $\Delta M=M_{N_5}-M_{N_4}=\Delta m_\nu$.
Therefore, besides the heavy-neutrino mass suppression, one also has a near cancellation between the different CP parities in the heavy neutrino pair. 
This is to be expected from the Quasi-Dirac nature of the heavy neutrino~\cite{Valle:1982yw}.  
Therefore, even for large mixing~($S_{e}\sim 10^{-2}$) and $M_{N}\sim\mathcal{O}(100\text{ GeV})$, the heavy neutrino contribution is totally negligible, about $10^{-12}$ smaller than the standard one.

\begin{figure}[!h]
\centering
\includegraphics[width=0.35\linewidth]{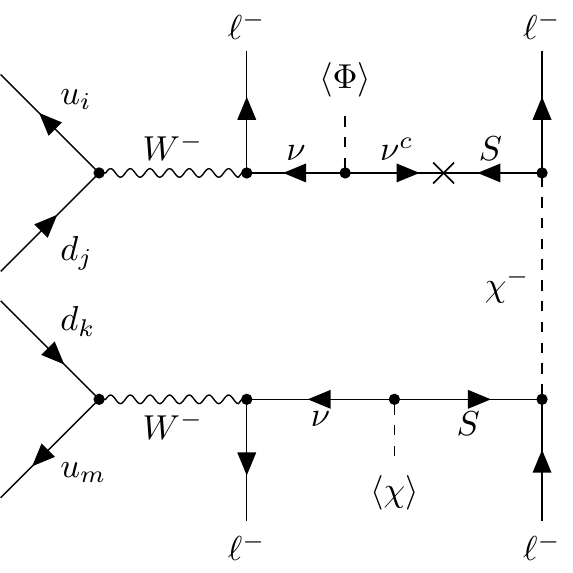}
\caption{\footnotesize{Feynman amplitudes for LNV tetraquark decays. There are similar $0\nu\beta\beta$-like and meson decay diagrams.}}
\label{fig:meson}
\end{figure}

The existence of high energy LNV processes $e^+ e^-\to \ell^+\ell^+ \bar{u}_i d_j \bar{u}_m d_k$ suggests higher-dimensional counterparts of the standard \znbb operator.
For example the operator $\bar{u}\bar{u}dd\bar{\ell}\bar{\ell}\bar{\ell}\ell$ is present in the vanilla type-I seesaw and in the inverse seesaw mechanism. 
However, the resulting LNV rates would be suppressed by the fourth power of the light-heavy neutrino mixing factor ${S^4_{e}}$.
In contrast, the contribution from our new operator is proportional to ${S^2_{e}} {Y^2_{S}}$.
Given that in the limit $v_\chi\to 0$ the Yukawa coupling $Y_{S}$ is hardly restricted, this represents a much milder suppression. Assuming crossing symmetry such operator would lead to a plethora of low-energy LNV observables, such as
\begin{align} 
& \textbf{$0\nu\beta\beta$-like decay: } d_j d_k \to u_i u_m \ell^-\ell^- \ell^+\ell^-, \\
& \textbf{Tetraquark decay: } \bar{u}_i d_j \bar{u}_m d_k \to \ell^-\ell^- \ell^+\ell^- ,\\
& \textbf{Meson decay: } M_1^- \to M_2^+ \ell^- \ell^- \ell^+\ell^-.
\end{align}
For example, the exotic LNV tetraquark decays are shown in Fig.~\ref{fig:meson}.

 There are also a number of LNV meson decays, discussed in recent papers~\cite{Dib:2016wge,Abada:2017jjx,Bolton:2019pcu,Chun:2019nwi}.
While these LNV decays are enhanced with respect to conventional seesaw expectations, they are probably still too small.
However, one might envisage other constructions where our new operator could have larger low-energy effects.
%%%%%%%%
\section{Conclusions}\label{sec:conclusions}
 In contrast to other low-scale seesaw schemes~\cite{Anamiati:2016uxp,Antusch:2017ebe,Drewes:2019byd,Fernandez-Martinez:2022gsu,Antusch:2022ceb,Antusch:2023nqd}, the linear seesaw mechanism provides a UV-complete setup where 
 LNV can be large at high energies, even at lepton colliders.
  In our scheme the small neutrino masses are sourced by the tiny VEV of a leptophilic scalar Higgs doublet. 
  The heavy Quasi-Dirac neutrino mediators and the new neutral and charged Higgs bosons can all be kinematically accessible to colliders. 
  Their production mechanisms are shown in Figs.~\ref{fig:HpN-production} and \ref{fig:feynman-production}. 
  Proposed lepton colliders such as ILC/CLIC/FCC-ee/CEPC or a muon collider now under discussion would provide ideal environments to probe such new physics. 
  Indeed, assuming a reasonable $\mathbf{Y}_S$ benchmark value we have verified that the cross sections for
$NN$ and $HH$ production at a 3~TeV $e^+ e^-$ collider can all be sizeable.  
Similarly, for the associated $NH$ production channel at a 3~TeV $e^- \gamma$ collider. 
Fig.~\ref{fig:CSeetoee4j} gives benchmark cross sections for $e^+e^-\to N \overline{N}\to \ell^\pm \ell^\pm 4j$~(left) and
$e^-\gamma\to H^-N\to \ell^-\ell^\pm \ell^\pm 4j$~(middle) versus the heavy neutrino mass $M_{N}$ assuming the heavy neutrino oscillation is effective $R_{\ell\ell}\approx 1$. 
The right panel shows the cross-section for the charged Higgs portal process $e^+ e^-\to H^+H^-\to$ $\ell^+\ell^-\ell^\pm \ell^\pm 4j$ versus the charged scalar mass. 
Similar results hold also for the $pp$ and a muon collider, as shown in the Fig.~\ref{fig:CSmumutoee4j}. 
 %                                                    %
One sees that LNV signatures are potentially large at high energies, in agreement with the black-box theorem, and with the suppression of their low-energy counterparts.
%%%%%%
\par In Sec.~\ref{sec:oscillation}, we have discussed the theory of heavy neutrino neutrino-to-antineutrino oscillation in order to shed light on the expected LNV/LNC rates.
We found that, while LNV involving virtual quasi-Dirac neutrino propagation is always suppressed by $\Delta M/M_N$, at high energies with on-shell heavy neutrinos, the expected suppression is only by $\Delta M/\Gamma_N$, and so the suppression can be mitigated. 
The key advantage of our linear seesaw model with respect to all other seesaw mechanisms, is that the primary heavy neutrino production rate is not neutrino-mass-suppressed, since it is determined by the $\mathbf{Y}_S$ Yukawa coupling instead of the small light-heavy neutrino mixing. This can lead to sizeable LNV rates even for $R_{\ell\ell}^{\rm obs}\ll 1$. 

 In addition, since our setup provides a very simple connection between the mass splitting between the two heavy neutrino mediators and the measured neutrino mass splittings in neutrino oscillations. As a result one has the potential to probe the neutrino mass ordering at colliders. 
Note also that large LNV at high energies, with small non-zero $v_\chi$, is consistent with low-energy limits, and the validity of the black-box theorem. 
Moreover, in the limit $v_\chi\to 0$ lepton number is conserved, but lepton flavour can be substantially violated.

\par All in all, heavy neutrino signatures from our leptophilic Higgs portal in the simplest linear seesaw provide a more promising neutrino probe than more complicated seesaws with extra gauge bosons.
Our model may be useful to shed light on the Majorana nature of neutrinos and help elucidate the role of lepton number and lepton flavour non-conservation.

%\newpage
\begin{center}
\bf Acknowledgments
\end{center}
The work of A.B. is supported by Funda\c{c}\~ao para a Ci\^encia e a Tecnologia (FCT, Portugal) through the PhD grant UI/BD/154391/2023 and
through the projects CFTP-FCT Unit UIDB/00777/2020 and UIDP/00777/2020, CERN/FIS-PAR/0019/2021, which are partially funded through POCTI (FEDER), COMPETE, QREN and EU.
  The work of S.M. is supported by KIAS Individual Grants (PG086002) at Korea Institute for Advanced Study.
  The work of RS and A.B. is supported by the Government of India, SERB Startup Grant SRG/2020/002303.
  J.W.F.V. is supported by the Spanish grants PID2020-113775GB-I00~(AEI/10.13039/501100011033) and Prometeo CIPROM/2021/054 (Generalitat Valenciana).
  
%%%%%%%%%%%%%%%%%%%%%%%%%%%%%
%%\end{acknowledgments}
%%%%%%%%%%%%%%%%%%%%%%%%%%%%%

\bibliographystyle{utphys}
\bibliography{bibliography} 
\end{document}